\documentclass[%
 reprint,
superscriptaddress,
 amsmath,amssymb,
 aps,
]{revtex4-2}

\usepackage{graphicx}
\usepackage{dcolumn}
\usepackage{bm}

\begin{document}

\title{Reaction-induced molecular dancing and boosted diffusion of enzymes}

\author{Hiroyuki Kitahata}
\email{kitahata@chiba-u.jp}
\affiliation{Department of Physics, Chiba university, Yayoi-cho 1-33, Inage-ku, Chiba 263-8522, Japan}

\author{Alexander S. Mikhailov}
\email{mikhailov@fhi-berlin.mpg.de}
\affiliation{Department of Physical Chemistry, Fritz Haber Institute of the Max Planck Society, Faradayweg 4 - 6, 14195 Berlin, Germany}
\affiliation{Computational Molecular Biophysics, WPI Nano Life Science Institute, Kanazawa University, Kakuma-machi, Kanazawa 920-1192, Japan} 

\begin{abstract}
A novel mechanism of reaction-induced active molecular motion, not involving any kind of self-propulsion, is proposed and analyzed. Because of the momentum exchange with the surrounding solvent, conformational transitions in mechano-chemical enzymes are accompanied by motions of their centers of mass. As we show, in combination with rotational diffusion,  such repeated reciprocal motions generate an additional random  walk - or molecular dancing - and hence boost translational diffusion of an enzyme.  A systematic theory of this phenomenon is developed, using as an example a simple enzyme model of a rigid two-state dumbbell. To support the analysis, numerical simulations are performed. Our conclusion is that the phenomenon of molecular dancing could underlie the observations of reaction-induced diffusion enhancement in enzymes.  Major experimental findings, such as the occurrence of leaps, the anti-chemotaxis, the linear dependence on the reaction turnover rate and on the rate of energy supply, become thus explained. Moreover, the dancing behavior is possible in other systems, natural and synthetic, too. In the future, interesting biotechnology applications may be developed using such effects.
\end{abstract}


\maketitle

\section{Introduction}

 When the phenomena of active motion on nano- and microscales have been previously considered, the discussion has been typically focused on self-propulsion mechanisms based on hydrodynamic interaction forces or on the chemotaxis, diffusio- and thermophoresis effects~\cite{dey-review,granick, ajdari,ajdari1,iima,sakaue,kapral-EPL,wolynes,gaspard}. Here, we bring attention to a different kind of reaction-induced active molecular motion where self-propulsion is not involved. 
 
 If a pair of human dancers steps forward and back along the same direction, they would keep their position on the floor. However, if the pair is \textit{waltzing}, i.e. turning between the steps, the entire floor would become explored. This is because a combination of reciprocal steps and arbitrary rotations generates a translational random walk. As we intend to show, similar dancing behavior is possible on molecular scales, leading to reaction-induced diffusion enhancement in enzymes.

Within its cycle, a mechanochemical enzyme undergoes conformational transitions induced by changes in its ligand state. Because of the momentum exchange between the enzyme and the surrounding fluid, such transitions are accompanied by shifts, or steps, in the spatial position of enzyme's center of mass (CM). The shifts in the CM position can be comparable to  enzyme's size.  Importantly, the CM shifts are almost reciprocal within the turnover cycles of real enzymes, so that they cancel one another and no significant self-propulsion occurs~\cite{wolynes,iima,sakaue}. 

In combination with rotational diffusion, the reciprocal CM shifts lead, however, to dancing of enzymes. The rotational diffusion on molecular scales is typically so fast that an enzyme would tumble after each next reaction event within its turnover cycle. Hence, an additional active random walk would arise, enhancing  translational diffusion of the enzyme. 

Boosted diffusion for catalytically active enzymes has been experimentally reported  by several research groups~\cite{butler,sengupta,dey,bustamante,granick, xu,granick-science,granick-master}. While this behavior has already been a subject of reviews~\cite{dey-review,gilson} and interesting biotechnology applications have been discussed~\cite{yuan}, its nature remains poorly understood. Importantly, it was demonstrated~\cite{wolynes} that the observed diffusion enhancement cannot be explained by self-propulsion due to hydrodynamic intramolecular interactions between moving protein domains, because it is too weak~\cite{ajdari,ajdari1,iima,sakaue}. Although other approaches have been more recently proposed~\cite{agudo,mandal, agudoC}, they are still only in partial agreement with the experimental results (see discussion in Section~\ref{sec:discussion} for the details). 

In studies of hydrodynamic molecular effects, simplified minimal models were previously used. Thus, for example, the three-bead Golestanian swimmer model~\cite{swimmer} was employed to analyze self-propulsion effects for enzymes~\cite{wolynes} and two-bead models were used to study diffusion enhancement for passive tracer particles in the solutions of active enzymes~\cite{kapralPNAS,dennison,kitahata,soft}. Moreover, the effects of hydrodynamic interactions between beads in an oscillating deformable dumbbell were analyzed  as a potential mechanism of diffusion enhancement for single enzymes~\cite{illien0,illien,kondrat,soft}. 

Below, we consider models of molecules formed by a set of beads, with conformations depending on the reaction state. In contrast to swimming, hydrodynamic interactions between the beads are not essential for the dancing behavior that we intend to explore. Because of this, the analysis is performed within the Langevin description where the action of fluid is taken into account merely through viscous friction forces applied to the beads and hydrodynamic interactions between them are dropped. 

The notions of the center of mass and of the rotation center of a molecule, that play an important role in our theory, are introduced and compared in Section~\ref{sec:com_rc}. To illustrate principal effects, a minimal model of a two-state rigid dumbbell is first chosen by us in Section~\ref{sec:dancing}.  We show that dancing generates a random walk where finite leaps are alternating with the intervals of smooth thermal Brownian motion. Later on, in Section~\ref{sec:general}, a general theory for enzymes consisting of an arbitrary number of interacting beads with various reactions, changing the conformations, is constructed. For clarity, only the summary of results is given there, while technical derivation details are moved to the Appendix. Numerical simulations, that confirm theoretical predictions, are presented in Section~\ref{sec:simulation}. The theory is compared with  experimental facts on boosted diffusion in catalytically active enzymes in Section~\ref{sec:discussion}. The conclusions are formulated in Section~\ref{sec:conclusion}.

\section{Center of mass \textit{vs.} rotation center \label{sec:com_rc}} 

For a molecule with $N$ beads of equal mass, the center of mass (CM) is defined as $\mathbf{R}=(1/N)\sum_i\mathbf{R}_i$. The rotation center (RC) of the molecule is given by $\mathbf{Q}=(1/\Gamma) \sum_i\gamma_i\mathbf{R}_i$.  Here, $\gamma_i=1/\mu_i$ is the friction coefficient of bead $i$, $\Gamma=\sum_i\gamma_i$ is the friction coefficient of the entire molecule, and  $\mu_i$ is the mobility of bead $i$. As shown in the Appendix, RC is a spatial position with respect to which the total torque vanishes when forces are applied to the beads in such a way that only translational motion of the entire molecule takes place. 

 The difference between the behavior of RC and CM can be illustrated by an example. Consider a dimer made by two interacting beads. In the overdamped Langevin description, the equations of motion for the beads are 
\begin{align}
    \gamma_1\frac{d\mathbf{R}_1}{dt}=\mathbf{f}+\boldsymbol{\xi}_1(t), \qquad \gamma_2\frac{d\mathbf{R}_2}{dt}=-\mathbf{f}+\boldsymbol{\xi}_2(t) \label{LanvevinEq}
\end{align}
 where $\mathbf{f}=-\partial U/\partial \mathbf{R}_1$ is the interaction force that depends on the distance $r_{12} = |\mathbf{R}_1-\mathbf{R}_2|$ between the beads,  and $\boldsymbol{\xi}_{1,2}(t)$ are independent thermal noises, such that 
 \begin{align}
 \langle \xi_{i,\alpha}(t)\xi_{j,\beta}(t')\rangle=2\gamma_i k_BT\delta_{ij}\delta_{\alpha\beta}\delta(t-t')    
 \end{align}
 where $(i,j)=1,2$ and $(\alpha,\beta)=(x,y,z)$.
 
  The equations of motion for the CM position $\mathbf{R}=(\mathbf{R}_1+\mathbf{R}_2)/2$ and for the deviation $\mathbf{r}_1=\mathbf{R}_1-\mathbf{R}$ from it are 
\begin{align}\label{2CM}
  \frac{d\mathbf{R}}{dt}&=\frac{1}{2}(\mu_1-\mu_2)\mathbf{f}+\boldsymbol{\chi}(t) , \nonumber \\
  \frac{d\mathbf{r}_1}{dt}&=\frac{1}{2}(\mu_1+\mu_2)\mathbf{f}+\boldsymbol{\rho}(t),
 \end{align}
where thermal noises are $\boldsymbol{\chi}(t)=(1/2)\big(\mu_1{\boldsymbol \xi}_1(t)+ \mu_2{\boldsymbol \xi}_2(t)\big)$ and $\boldsymbol{\rho}(t)=(1/2)\big(\mu_1{\boldsymbol \xi}_1(t)- \mu_2{\boldsymbol \xi}_2(t)\big)$. Note that we have
\begin{align}
    \langle \chi_\alpha (t) \rho_\beta (t')\rangle=\frac{1}{2}(\mu_1-\mu_2)k_BT\delta_{\alpha\beta}\delta(t-t'). 
\end{align}
 
 If the mobilities of beads are different ($\mu_1\neq \mu_2$), the motion of CM depends through the interaction force $\mathbf{f}$ on the dimer orientation and the distance between the beads. This means that the CM position is not conserved: it is sensitive to the conformation and orientation of the molecule. Additionally, a dependence of the CM position on dimer's conformation and orientation arises through correlations between thermal noises $\boldsymbol{\chi}(t)$ and $\boldsymbol{\rho}(t)$.
 
 On the other hand, the equations of motion for the RC of the dimer, $\mathbf{Q}=(\gamma_1\mathbf{R}_1+\gamma_2\mathbf{R}_2)/\Gamma$, and for the deviation $\mathbf{q}_1=\mathbf{R}_1-\mathbf{Q}$ from it are
 \begin{align}
     \Gamma\frac{d\mathbf{Q}}{dt}={\boldsymbol \eta}(t), \qquad \gamma_1\frac{d\mathbf{q}_1}{dt}=\mathbf{f}+\boldsymbol{\zeta}(t),
 \end{align}
where $\Gamma = \gamma_1+\gamma_2$. For thermal noises ${\boldsymbol \eta}(t)=\boldsymbol{\xi}_1(t)+\boldsymbol{\xi}_2(t)$ and $\boldsymbol{\zeta}(t)=(1/\Gamma)(\gamma_2\boldsymbol{\xi}_1(t)-\gamma_1\boldsymbol{\xi}_2(t))$, we have 
 \begin{align}
   \langle \eta_\alpha(t)\eta_\beta(t')\rangle=2\Gamma k_BT \delta_{\alpha\beta}\delta(t-t'),
    \end{align}
  \begin{align}  
   \langle \zeta_\alpha(t)\zeta_\beta(t')\rangle=2(\gamma_1\gamma_2/\Gamma)k_BT \delta_{\alpha\beta}\delta(t-t').  
 \end{align}
 Importantly, these noises are independent, i.e. 
 \begin{align}
     \langle \eta_\alpha(t)\zeta_\beta (t')\rangle=0.
 \end{align}

Hence, the rotation center $\mathbf{Q}(t)$ performs free Brownian motion statistically independent from conformational and orientational fluctuations described by the Langevin equation for $\mathbf{q}_1(t)$. This means that, in contrast to CM,  the motion of RC is separate from the instantaneous conformation and orientation of the dimer. 

As shown further in Appendix, such separation of motions holds generally for a molecule with $N$ beads with arbitrary interactions between them. The motion of CM is affected by conformational fluctuations of the molecule, whereas the motion of RC is independent from them. Thermal rotational fluctuations take place around RC, but not around CM. 

Note that shifts of CM in response to intramolecular conformational transitions and to molecular rotations manifest an exchange of mechanical momentum between the molecule and the surrounding fluid. Although sometimes overlooked, such momentum exchange is actually  present even within the Langevin description where hydrodynamic  interactions between the particles (such as in the Oseen approximation) are completely dropped. Fluid dynamics enters into the description only through the viscous friction force.

\section{Dancing of dumbbells \label{sec:dancing}}
In this section, the phenomena of molecular dancing will be introduced by using, as an example, a minimal enzyme model of a  reactive dumbbell.
\subsection{The reactive dumbbell model}

\begin{figure}
\centering
\includegraphics{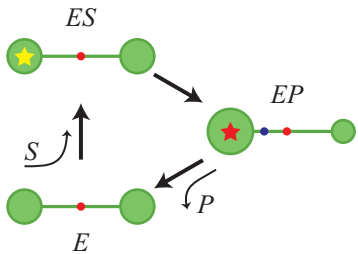}
\caption{Reaction cycle of the model dumbbell enzyme. The dumbbell consists of two beads connected by a link of a fixed length $2a$. Substrate $S$ binds in the state $E$ to domain 1 (left) and the substrate-enzyme complex $ES$ is formed. In this state, rapid conversion of substrate into product $P$ occurs, so that a transition to the product-enzyme complex  $EP$ takes place. After that, the product is released. The distance between the beads remains fixed, but the viscous friction coefficients of the beads become changed in the state $EP$. The RC and CM of the dumbbell are shown by red and blue dots, respectively. The two centers coincide in the states $E$ and $ES$\, but they are shifted one from another in the state $EP$}
\label{fig:Fig1}
\end{figure}

The model dumbbell enzyme consists of two beads of equal mass connected by a stiff link of length $2a$. For real enzymes, the beads would have corresponded to two protein domains. Only domain 1 is catalytically active: substrate $S$ can bind to it, become converted to product $P$ and then be released. Hence, the reaction is (Fig.~\ref{fig:Fig1})
\begin{align}
    S+E\rightarrow ES \rightarrow EP \rightarrow E+P
\end{align}
and instantaneous evacuation of the product takes place. Note that the irreversibility of this reaction implies a large difference in the Gibbs free energy between the substrate and the product.  

On the considered timescales, inertia is negligible and the overdamped Langevin description holds. Hence, the beads are characterized by their mobility $\mu_{1,2}$ and viscous friction coefficients $\gamma_{1,2}=1/\mu_{1,2}$. The friction coefficients of the beads depend on the chemical state. We shall assume that $\gamma_1=\gamma_2=\gamma$ in the states \textit{E} and \textit{ES}, but $\gamma_1=\gamma +\Delta\gamma$ and $\gamma_2=\gamma-\Delta\gamma$ in the state \textit{EP}. Hence, bead 1 becomes less mobile and bead 2 is more mobile when the product is formed.

It will be further assumed that the substrate-enzyme complex \textit{ES} has a very short lifetime, so that it becomes immediately converted into the product-enzyme complex \textit{EP}. Hence, in the final employed  model, there are only two states: $s=0$ (\textit{E}) and $s=1$ (\textit{EP}), and the reaction scheme is reduced to 
\begin{align}
    S+E \xrightarrow{\kappa^{10}} EP \xrightarrow{\kappa^{01}} E+P
\end{align}
where $\kappa^{10},\kappa^{01}$ are reaction rate constants and $\kappa^{10}=\nu c_s$ (with $c_s$ being the substrate concentration). 

For the dumbbell, the CM always lies in its geometric center. The RC is shifted from the geometric center by distance
\begin{align}\label{shift0}
 d=\frac{\Delta\gamma}{\gamma}a   
\end{align}
in the state $EP$ ($s=1$) and coincides with it in the state $E$ ($s=0$); here $a$ is the half-length of the dumbbell. Hence, chemical transitions between the states within a turnover cycle are accompanied by RC shifts.

The reactive dumbbell model can be viewed as an idealization for an enzyme protein  with two domains. The friction coefficient of a protein domain can be written as $\gamma=6\pi \nu R$, where $\nu$ is the solvent viscosity, thus introducing the effective hydrodynamic (or Stokes) radius $R$ of the domain. The hydrodynamic radius of a protein represents the radius of a spherical particle that, within the Stokes approximation, would have the same friction coefficient as it. Importantly, the hydrodynamic radius can change even though the mass of a protein remains conserved. For example, in a folding-unfolding transition, the unfolded protein shall have a larger hydrodynamic radius than the folded one. Any conformational change within a protein domain generally affects its friction coefficient and its hydrodynamic radius. It should be stressed that the model does not assume the mass transfer between the two domains.

\subsection{Reaction-induced random walk} 
\begin{figure}
\centering
\includegraphics{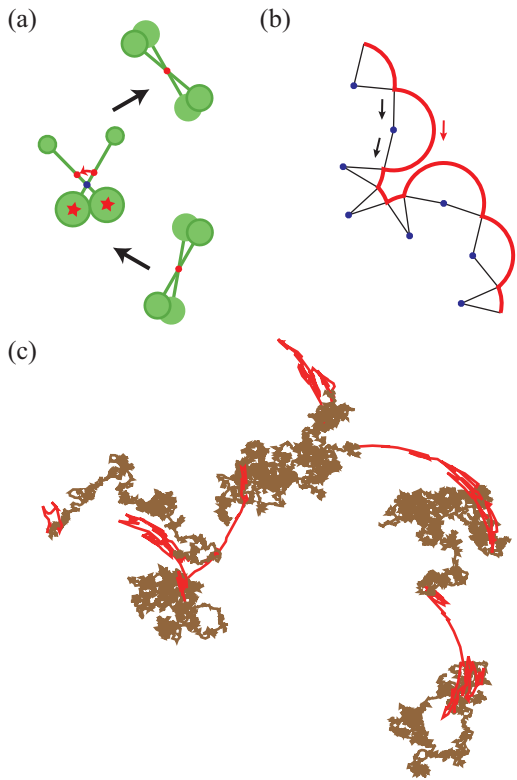}
\caption{(a) In the state $EP$, the RC (blue dot) of the dumbbell is shifted away from its CM (red dot), whereas in the state $E$  both centers coincide. Rotational diffusion induces CM rotation (along the red circular segment) in the state $EP$, but has no effect on CM in the state $E$. (b) Rotations of CM around various centers in consequent reaction cycles generate a random walk. Trajectories of CM (red) and RC (black) are shown. RC shifts occur under transitions $E\rightarrow EP$ and $EP\rightarrow E$, and CM rotations take place in the states $EP$. Thermal Brownian motion is, for simplicity, dropped in this plot. (c) A typical CM trajectory (schematic) for the dumbbell. Intervals of thermal Brownian motion (brown) alternate with random leaps (red).}
\label{fig:Fig2}
\end{figure}

 In chemical transitions $E\rightarrow EP$ and $EP\rightarrow E$, RC of the dumbbell moves away from CM and back towards it. As has been shown in Section~\ref{sec:com_rc}, rotational diffusion of the dumbbell always takes place around RC. Because CM is shifted from RC in the state $EP$ ($s=1$), it would therefore rotate around RC under thermal orientational fluctuations in this state (Fig.~\ref{fig:Fig2}a). On the other hand, because CM coincides with RC in the state $E$ ($s=0$), its position would not be changed by rotational diffusion in this state. 

 If only rotational diffusion is taken into account, the trajectories of CM and RC would therefore look as shown in Fig.~\ref{fig:Fig2}b. The RC makes jumps of equal length $d$ along the instantaneous dumbbell directions in chemical transitions, but remains immobile in the intervals between them.  The CM position is not changed under such transitions. However, CM makes random rotations around RC due to thermal rotational diffusion of the dumbbell when it is in the state $s=1$.
 
If translational diffusion is additionally included,  the CM would thus perform a random walk, i.e. molecular dancing, that is a superposition of thermal Brownian motion and rapid leaps (Fig.~\ref{fig:Fig2}c). The leaps of CM represent random thermal rotations of CM around RC in the chemical state $EP$; they occur under the reaction and are of molecular size.

\subsection{Diffusion enhancement}
The leaps would contribute to the molecular mean-square displacement over a long time and, under the reaction, dumbbell diffusion would become enhanced. The additional contribution to the diffusion coefficient, arising because of the molecular dancing effect, can be estimated in the following way.

Within the lifetime $\tau_1$ of the state $s=1$, the dumbbell would change due to rotational diffusion its orientation from $\mathbf{n}=\mathbf{n}(t)$ to $\mathbf{n}'=\mathbf{n}(t+\tau_1)$. Because CM is shifted by distance $d$ from the center around which random rotation takes place, its spatial position would become shifted by vector $\mathbf{u}=d(\mathbf{n}'-\mathbf{n})$ in each turnover cycle.

For the reaction-induced random walk, the mean-square CM displacement over the time $t_0$ comprising many independent turnover cycles can be written in the three-dimensional case as $\langle \Delta\mathbf{R}(t_0)^2\rangle=\langle \mathbf{u}^2\rangle kt_0$ where $k$ is the turnover rate (so that $kt_0$ gives the mean number of reaction cycles within time $t_0$). Because the diffusion coefficient for a random walk can generally be determined as $D=(1/6)\lim_{t_0\rightarrow \infty}(1/t_0)\langle \Delta\mathbf{R}(t_0)^2\rangle$, reaction-induced dancing increases the diffusion coefficient by $D_\text{dance}= (1/6)k\langle \mathbf{u}^2\rangle$. 

The mean-square leap $\mathbf{u}(\tau_1)=d\big(\mathbf{n}(t+\tau_1)-\mathbf{n}(t)\big)$ within the lifetime time $\tau_1$ of the state $EP$ is
\begin{align}
  \langle\mathbf{u}(\tau_1)^2\rangle=&2d^2 \left[1-\langle \mathbf{n}(t+\tau_1)\mathbf{n}(t)\rangle \right] \nonumber \\
  =& 2d^2\left[1-\exp(-D^1_\mathrm{rot}\tau_1)\right]  
\end{align}
where $D^1_\mathrm{rot}$ is the rotational diffusion coefficient of the dumbbell in the state \textit{EP}. 

The lifetimes $\tau_1$ of the state $EP$ have probability distribution $p(\tau_1)=\kappa^{01}\exp{(-\kappa^{01}\tau_1)}$. After taking the average over $\tau_1$, we determine a contribution to the diffusion coefficient due to molecular dancing effect
\begin{align}\label{enhancement}
    D_{\mathrm{dance}}=\frac{1}{3}kd^2\frac{D^1_\textrm{rot}}{\kappa^{01}+D^1_\mathrm{rot}}.
\end{align}

In the derivation of equation (\ref{enhancement}), we have assumed that each next reaction cycle starts from a random and independent orientation of the dumbbell. This assumption is satisfied if the molecule would typically tumble many times between the cycles. In other words, the orientational correlation time $\tau^0_\mathrm{rot}=1/D^0_\mathrm{rot}$ in the ligand-free state $s=0$ must be much shorter than the waiting time $\tau_0=1/\kappa^{10}$ for substrate binding and, thus, initiation of the next cycle. Hence, the condition $D^0_{\mathrm{rot}}\gg \kappa^{10}$ must be satisfied.

The result (\ref{enhancement}) clearly shows that the considered diffusion enhancement emerges from an interplay between chemical transitions and rotational diffusion. Indeed, it can be seen that $D_{\mathrm{dance}}$ vanishes in the limit of $\kappa^{01}\gg D^1_\mathrm{rot}$, i.e. when the lifetime of the state $s=1$ is so short that the dumbbell does not significantly change its orientation within it. In the opposite limit of very fast rotational diffusion ($\kappa^{01} \ll D^1_\mathrm{rot}$), the dependence on $D^1_\mathrm{rot}$ disappears and we have a simple result 
\begin{align}
 D_{\mathrm{dance}}= \frac{1}{3}kd^2.   
\end{align}

Generally, diffusion constants of the dumbbell are different in the states $s=0$ and $s=1$. If the viscous friction coefficients of the dumbbell in these two states are $\Gamma_0$ and $\Gamma_1$, the respective diffusion constants are $D^0=k_BT/\Gamma_0$ and $D^1=k_BT/\Gamma_1$. If $\pi^0$ and $\pi^1$ are occupation probabilities of the two states (that is, relative times spent by the molecule in each of them), the total effective diffusion coefficient for the reactive dumbbell can therefore be estimated as $D_{\mathrm{eff}}=D^0\pi^0+D^1\pi^1+D_{\mathrm{dance}}$. 

In the considered example, where $\gamma_1=\gamma_2=\gamma$ in the state $s=0$ and $\gamma_1=\gamma +\Delta\gamma$ and $\gamma_2=\gamma-\Delta\gamma$ in the state $s=1$, we have however $D^1=D^0$ and, therefore, $D_{\mathrm{eff}}=D^0+D_{\mathrm{dance}}$. Hence, the contribution from molecular dancing yields the complete reaction-induced diffusion enhancement in this special case, i.e.  $\Delta D=D_\mathrm{eff}-D^0=D_{\mathrm{dance}}$.

Note that the CM random walk under molecular dancing can be approximated by a classical diffusion process only on the length and timescales longer than the size and duration of an CM leap, i.e. on the length scale larger than $d$ and the timescale larger than the characteristic lifetime $1/\kappa^{01}$ of the state $s=1$.

\subsection{Anti-chemotaxis for dumbbells}
Now, we consider a situation when a spatial gradient of the substrate concentration $c_s$ is present for considered dumbbells. 

The evolution equation for the probability density $ P(\mathbf{R},t)$ of the CM position $\mathbf{R}$ of the dumbbells can be derived via the Kramers-Moyal expansion \cite{risken}. Generally, this Fokker-Planck  equation has the form
\begin{align}
    \frac{\partial P}{\partial t}=-\frac{\partial}{\partial R_\alpha}\left[K^{(1)}_\alpha(\mathbf{R})P\right]+\frac{1}{2}\frac{\partial^2}{\partial R_\alpha\partial R_\beta}\left[K^{(2)}_{\alpha\beta}(\mathbf{R})P\right]
\end{align}
where $\{\alpha,\beta\}=\{x,y,z\}$ and summation over the repeated indices is assumed. The coefficients are \begin{align}
K_\alpha^{(1)}(\mathbf{R})=\lim_{\Delta t\rightarrow\infty} \frac{\langle \Delta R_\alpha(\Delta t) \rangle}{\Delta t},     
\end{align}
 \begin{align}
   K_{\alpha\beta}^{(2)}(\mathbf{R})=\lim_{\Delta t\rightarrow\infty} \frac{\langle \Delta R_\alpha(\Delta t)\Delta R_\beta(\Delta t) \rangle}{\Delta t}  
 \end{align}
 where $\Delta R_\alpha(\Delta t)=R_\alpha(t+\Delta t)-R_\alpha(t)$. 

We assume that rotational diffusion is fast, i.e. $D^0_\mathrm{rot}\gg k$. Then, the dumbbell would tumble many times between its next cycles, with all its orientations becoming equally probable. Hence, even if a spatial gradient of substrate concentration is present, it cannot impose a preferred molecular direction of motion. The resulting symmetry implies that $K^{(1)}_\alpha=0$ and that $K^{(2)}_{\alpha\beta}=2 D_{\mathrm{eff}} \delta_{\alpha\beta}$ where $D_\mathrm{eff}(\mathbf{R})=D^0+D_\mathrm{dance}(\mathbf{R})$ is the local diffusion coefficient. 
Thus, the evolution equation for the local enzyme concentration $c=N_0 P$ will be
\begin{align}\label{anti0}
    \frac{\partial c}{\partial t}=\nabla^2\left[D_{\mathrm{eff}} (\mathbf{R})c\right].
\end{align}
where $N_0$ is the total number of enzymes.

For low substrate concentrations $c_s$, the turnover rate $k$ in equation (\ref{enhancement}) is proportional to substrate concentration $c_s$, implying that the effective diffusion coefficient is also linear in $c_s$, i.e. 
$ D_{\mathrm{eff}}(\mathbf{R})=D^0+\lambda c_s(\mathbf{R})$ with some constant coefficient $\lambda$. Substituting this into the above equation and transforming it to the standard form of the diffusion equation, we obtain
\begin{align}
    \frac{\partial c}{\partial t}=\nabla\left[\left(\lambda\nabla c_s\right) c\right]+\nabla \left(D_\mathrm{eff} \nabla c \right).
\end{align}
Hence, a spatial gradient in the substrate concentration induces the flow $\mathbf{J}=-\lambda \nabla c_s$ of dumbbells in the direction opposite to the substrate concentration gradient $\nabla c_s$. 

\section{General theory of molecular dancing \label{sec:general}}
In this section, a systematic theory of molecular dancing will be developed. For clarity, only a summary of results will be presented, while technical derivation details shall be given separately in the Appendix.

\subsection{The model}

The model enzyme molecule consists of $i=1,2,..,N$ beads of equal mass. It can be found in different chemical states $s$. The beads $i$ and $j$ interact via pair interaction potentials $U^s_{ij}(r_{ij})$ that are functions of the distances $r_{ij}= \left| \mathbf{R}_i-\mathbf{R}_j \right|$ between them, where $\mathbf{R}_i$ denotes the position of the bead $i$. The interaction potentials generally depend on the chemical state $s$. In the Langevin description, each bead $i$ is characterized by its mobility coefficient $\mu_i^s$ depending on the chemical state $s$. 

Reactions represent stochastic transitions between chemical states. To simplify derivations,  we assume that each reaction $s'\rightarrow s$ occurs only in a definite conformation determined by a certain set of distances  $\{r_{ij}^{ss'}\}$ between the beads. Hence, conformation-dependent reaction rate constants are
\begin{align}\label{rates}
    W^{ss'}=w^{ss'}\prod_{i,j}\delta \left(\left|\mathbf{R}_i-\mathbf{R}_j\right|-r_{ij}^{ss'}\right),
\end{align}
where the delta-functions take into account that a reaction $s'\rightarrow s$ is possible only in a definite configuration.

\subsection{Master equations}

The system is characterized by a set of multi-particle probability distributions $p^s(\{\mathbf{R}_k\},t)$ in different chemical states. They satisfy a system of coupled master equations
\begin{align}
\label{master}
    \frac{\partial p^s}{\partial t}=-\sum_{i}\frac{\partial j^s_{i,\alpha}}{\partial R_{i,\alpha}}+ \sum_{s'}\left(W^{ss'}p^{s'}-W^{s's}p^{s}\right).
\end{align}
Here and below, summation over repeated Greek indices $\alpha=\{x,y,z\}$ is always assumed. In the overdamped Langevin description, probability flows are 
\begin{align}
    j^s_{i,\alpha}= -\mu^s_i\left(\sum_{j}\frac{\partial U^s_{ij}(r_{ij})}{\partial R_{i,\alpha}}p^s+k_BT\frac{\partial p^s}{\partial R_{i,\alpha}}\right),
\end{align}
where the last term accounts for thermal noise. 
 
 
 On the other hand, joint probability distributions $p^s_{RC}(\mathbf{Q}^s,\{\mathbf{q}^s_k\},t)$ for the RC positions $\mathbf{Q}^s$ and for the deviations from them $\mathbf{q}^s_k=\mathbf{R}_k-\mathbf{Q}^s$ in different chemical states obey the following evolution equations (see Appendix for the derivation):
\begin{align}\label{masterRC}
 &\frac{\partial p^s_{RC}}{\partial t} =-\frac{\partial H^s_{\alpha}}{\partial Q^s_{\alpha}}- \sum_i\frac{\partial h^s_{i,\alpha}}{\partial q^s_{i,\alpha}} \nonumber \\
 &+ \sum_{s'}\left[ w^{ss'}\prod_{i,j} \delta\left(\left|\mathbf{q}^{s}_i-\mathbf{q}^{s}_j\right| -r_{ij}^{ss'}\right) \right. \nonumber \\
& \left. \qquad \qquad \times  p^{s'}_{RC} \left(\mathbf{Q}^s-\mathbf{d}^{ss'}, \{\mathbf{q}^s_k+\mathbf{d}^{ss'}\}\right)\right. \nonumber \\
 & \left.\qquad -w^{s's}\prod_{i,j} \delta \left(\left|\mathbf{q}^{s}_i-\mathbf{q}^{s}_j\right| -r_{ij}^{s's}\right) p^s_{RC}\left(\mathbf{Q}^s,\{\mathbf{q}^s_k\}\right)\right].
 \end{align}
Here, $\mathbf{d}^{ss'}=\mathbf{Q}^s-\mathbf{Q}^{s'}$ is the shift in the RC position in the reaction $s'\rightarrow s$; it is given by equation (\ref{change}). The probability fluxes $\mathbf{H}^s$ and $\mathbf{h}^s_i$ are given by equations (\ref{driftH}) and (\ref{eqnA}) in the Appendix.

A remarkable property of evolution equations (\ref{masterRC}) is that, in absence of reactions, the RC dynamics is decoupled from conformational dynamics and from rotations of the enzyme molecule.

\subsection{Effective diffusion}
 
 As shown in the Appendix, starting from the master equation (\ref{masterRC}), an approximate closed evolution equation for the CM probability distribution $P(\mathbf{R},t)$ on long time and length scales can be derived assuming that locally the reaction is in a steady state. For local enzyme concentration $c = N_0 P$, an approximate diffusion equation is thus obtained, with the effective diffusion constant given by 
\begin{align}\label{eff}
    D_{\mathrm{eff}}=\sum_s\bar{\pi}^s D^s+\frac{1}{6}\sum_{s,s'}k^{ss'}\left(d^{ss'}\right)^2.
\end{align}
Here, $k^{ss'}$ is the mean rate, given by equation (\ref{rateseff}), of the reaction $s'\rightarrow s$ in the steady state , $\bar{\pi}^s$ is the respective occupation probability of the chemical state $s$, $d^{ss'}$ is the length of a RC shift accompanying a chemical transition from $s'$ to $s$, 
and $D^s=k_BTM^s$ is the equilibrium diffusion constant of the molecule in the conformation corresponding to the state $s$. To determine diffusion enhancement, one needs to subtract from $D_{\mathrm{eff}}$ the diffusion constant of the molecule in absence of reactions.

As an example, an irreversible reaction cycle
\begin{align}\label{cycle1}
s_0 \xrightarrow{\kappa^{10}} s_1 \xrightarrow{\kappa^{21}} s_2 \rightarrow  \cdots \rightarrow s_K \xrightarrow{\kappa^{0K}} s_0
\end{align}
can be considered. In this case, the rates of all transitions in the steady state are the same and equal to the turnover rate $k$. Hence, diffusion enhancement is given by
\begin{align}\label{deltaD}
    \Delta D=\sum_{l=1,\dots ,K} \bar{\pi}^l \left(D^l - D^0 \right)+\frac{1}{6}k\sum_{l=0,\dots ,K}\left(d^{l+1,l}\right)^2
\end{align}
where the notation $d^{K+1,K}=d^{0,K}$ is used and $D^0$ is the diffusion constant in the absence of the reaction (i.e., in the state with $l=0$). The first term takes into account a change in the diffusion constant because new states can become occupied under the reaction and the second term is due to the molecular dancing effect. 

For the model dumbbell enzyme with two states $s_0$ and $s_1$ and the reaction $s_0\rightarrow s_1\rightarrow s_0$, where the first reaction step corresponds to binding of substrate, we have
\begin{align}\label{two}
    \Delta D= \left(D^1 - D^0\right)\bar{\pi}^1+\frac{1}{3}kd^2 
\end{align}
with $d=d^{10}=d^{01}$ and
\begin{align}
    \bar{\pi}^1=\frac{\kappa^{10}}{\kappa^{01} + \kappa^{10}}
= \frac{\nu_s c_s}{\kappa^{01}+\nu_s c_s},
\end{align}
\begin{align}
 k= \dfrac{\kappa^{10}\kappa^{01}}{\kappa^{01} + \kappa^{10}}=\nu_s c_s \left(1 - \bar{\pi}^1\right),  
\end{align}
where $\kappa^{10}=\nu_s c_s$ is the substrate binding rate constant, $c_s$ is the substrate concentration and $\nu_s$ is a proportionality constant.

Note that the first term in equation (\ref{two}) coincides with the estimate for the diffusion enhancement in two-state enzymes previously obtained in Refs. \cite{agudo,mandal, agudoC}. The second term arises due to the effect of molecular dancing introduced by us.

The approximate diffusion description holds on the length scales longer than the characteristic size $d^{ss'}$ of a leap and on the timescales larger than the typical duration of a leap.

\subsection{Equilibrium systems}
At thermal equilibrium, transition rate constants $\kappa^{ss'}$ satisfy the conditions of detailed balance. The rates of forward and reverse reactions are equal for each reaction step (i.e. $k^{ss'}=k^{s's}$). The chemical potentials of the substrate and the product are the same. The occupation probabilities $\pi^s$ of different states $s$ are determined only by the Gibbs free energies of these states and do not depend on the reaction rates

Since deviations from equilibrium were not explicitly assumed, the developed theory of molecular dancing is applicable to equilibrium systems too. The equation (\ref{eff}) for the effective diffusion constant $D_\mathrm{eff}$ holds in this case as well. A special feature of equilibrium systems is that, because $k^{ss'}=k^{s's}$ at equilibrium and because $\mathbf{d}^{ss'}=-\mathbf{d}^{s's}$, the contributions into the diffusion from forward and reverse reactions are equal at each elementary reaction step. 

As an illustration, a chain of equilibrium reversible reactions can be considered,
\begin{align}\label{chain}
s_0 \overset{\kappa^{10}}{\underset{\kappa^{01}}{\rightleftarrows}} s_1 \overset{\kappa^{21}}{\underset{\kappa^{12}}{\rightleftarrows}} s_2 \rightleftarrows  \cdots \rightleftarrows s_K \overset{\kappa^{0K}}{\underset{\kappa^{K0}}{\rightleftarrows}} s_0.
\end{align}
Here, the first forward transition corresponds to binding of substrate and the last forward transition represents product release. For the rate constants of forward substrate and reverse product binding, we have $\kappa^{10}=\nu_sc_s$ and $\kappa^{K0}=\nu_pc_p$ where $c_s$ and $c_p$ are substrate and product concentrations, and $\nu_s$ and $\nu_p$ are proportionality coefficients. The detailed balance implies that, for each pair of states $s^l$ and $s^{l+1}$, we have
\begin{align}
\kappa^{l,l+1}=\kappa^{l+1,l}\exp{\left[-\Big(\mathcal{E}^l-\mathcal{E}^{l+1}\Big)/k_BT\right]}
\end{align}
where $\mathcal{E}^l$ is the Gibbs free energy of the state $s_l$. Moreover, 
\begin{align}
\kappa^{01}=\nu_sc_s\exp{\left[-\Big(\mathcal{E}^0-\mathcal{E}^{1}\Big)/k_BT\right]}
\end{align}
and
\begin{align}
\nu_pc_p=\kappa^{0K}\exp{\left[-\Big(\mathcal{E}^K-\mathcal{E}^0\Big)/k_BT\right]}.
\end{align}
As follows from the above equations, the equilibrium is reached at the product concentration $c_p$ satisfying the equation
\begin{align}
\nu_pc_p=\nu_sc_s\exp{\left[-\Big(\mathcal{E}^0-\mathcal{E}^K\Big)/k_BT\right]}.
\end{align}
It means that the chemical potentials of the substrate and the product become equal once the equilibrium is achieved.

The effective diffusion constant for the enzymes catalyzing the equilibrium reaction (\ref{chain}) is
\begin{align}\label{deq}
    D_\mathrm{eff}=\sum_{l=0,\dots ,K} D^l\bar{\pi}^l +\frac{1}{3}\sum_{l=0,\dots ,K}\kappa^{l+1,l}\bar{\pi}^l\left(d^{l+1,l}\right)^2
\end{align}
where we have again used the notations $\kappa^{K+1,K}=\kappa^{0K}$ and $d^{K+1,K}=d^{0K}$. The equilibrium occupation probabilities $\bar{\pi}^l$ of different states $l=0,\dots ,K$ are
\begin{align}
\bar{\pi}^l=\frac{\exp{\Big(-\mathcal{E}^l/k_BT\Big)}}{\sum_{l=0,\dots ,K}\exp{\Big(-\mathcal{E}^l/k_BT\Big)}}.
\end{align}

The first term in equation (\ref{deq}) represents the weighted average of diffusion coefficients in absence of reactions in different states; it does not depend on reaction rates. The second term is due to the dancing effects. As we see, each elementary equilibrium reaction gives a contribution to the diffusion coefficient proportional to its rate. Because contributions from forward and reverse reaction steps are equal, summation is left only over the forward reactions, with the factor 1/6 changed to 1/3.

Under reaction-induced diffusion enhancement, only a kinetic coefficient (i.e., the diffusion constant controlling the rate of relaxation for concentration distribution of enzymes) is increased. Hence, diffusion boosting does not imply that work has to be performed. Therefore, diffusion enhancement is possible under reversible equilibrium chemical reactions too. This has been pointed out in earlier publications \cite{illien0,illien} and it also follows from the theory constructed above.

Generally, it is well-known that any reaction generates additional \textit{internal} noise, whose intensity is proportionall to the respective reaction rate (see ref.\cite{gardiner,mikhailovI}). However, only the effects of reaction noise on fluctuations of chemical concentrations have been considered so far. As found in the present study, reaction noise can also lead to diffusion enhancement through the molecular dancing mechanism.

Although reaction-induced diffusion enhancement in equilibrium systems is principally important, this effect would typically be much weaker than diffusion boosting in non-equilibrium and practically irreversible reactions with high energy release.This is because, at equilibrium, only the states $s$ with the energies $\mathcal{E}^s$, differing by about the thermal energy $k_BT$, could be occupied. Usually, molecular conformations would not differ much over the states that could be reached by thermal fluctuations, so that the RC shifts $\mathbf{d}^{ss'}$in transitions between them would be small.

\subsection{Systems with spatial gradients}
Systems with spatial gradients can be furthermore considered. Both the solution parameters and the reaction rates may depend on the spatial position $\mathbf{R}$, leading to spatial dependence of $D^s$, $k^{ss'}$ and $\bar{\pi}^s$. For such systems, an approximate evolution equation for distribution of enzymes is derived in Appendix. 

Written in the standard form of a diffusion equation with a drift term, the evolution equation (\ref{antii}) for the local enzyme concentration $c$ becomes
\begin{align}\label{anti2}
    \frac{\partial c}{\partial t}=-\frac{\partial}{\partial \mathbf{R}}\left[\mathbf{V}(\mathbf{R})c\right]+\frac{\partial}{\partial \mathbf{R}}\left[D_{\mathrm{eff}}(\mathbf{R})\frac{\partial c}{\partial \mathbf{R}}\right]
\end{align}
where
  \begin{align}\
      D_{\mathrm{eff}}(\mathbf{R})=\sum_sD^s(\mathbf{R})\bar{\pi}^s(\mathbf{R})+ \frac{1}{6}\sum_{s,s'}k^{ss'}(\mathbf{R})\left(d^{ss'}\right)^2
  \end{align}
and
\begin{align}\label{velocity}
    \mathbf{V}(\mathbf{R})=-\sum_sD^s(\mathbf{R})\frac{\partial \bar{\pi}^s}{\partial \mathbf{R}}-\frac{1}{6}\sum_{s,s'}\left(d^{ss'}\right)^2\frac{\partial k^{ss'}}{\partial \mathbf{R}}
\end{align}
with $k^{ss'}(\mathbf{R})=\kappa^{ss'}(\mathbf{R})\bar{\pi}^s(\mathbf{R})$. This general equation provides an opportunity to discuss drift and diffusion under different kinds of spatial gradients. 

Suppose first that only the diffusion coefficients $D^s$ in various chemical states depend on the spatial position, whereas the reaction rates $k^{ss'}$ and the occupation probabilities $\bar{\pi}^s(\mathbf{R})$ of chemical states are independent of $\mathbf{R}$. As follows from equation (\ref{velocity}), we have $\mathbf{V}(\mathbf{R})=0$ in this case. Therefore, the evolution equation (\ref{anti2}) is reduced to  
\begin{align}\label{classical}
    \frac{\partial P}{\partial t}=\frac{\partial}{\partial R_\alpha}\left[D_{\mathrm{eff}}(\mathbf{R})\frac{\partial c}{\partial R_\alpha}\right]
\end{align}
where
  \begin{align}\
      D_{\mathrm{eff}}(\mathbf{R})=\sum_sD^s(\mathbf{R})\bar{\pi}^s+ \frac{1}{6}\sum_{s,s'}k^{ss'}\left(d^{ss'}\right)^2.
\end{align}
Note that diffusion coefficients $D^s$ can have spatial variation because such parameters as temperature or solution viscosity depend on the particle position in space. 

Equation (\ref{classical}) is the classical diffusion equation in systems with spatial gradients; it does not include the drift. According to it, in the steady state in absence of flux, the particles should be \textit{uniformly} distributed, despite the presence of the gradients:
\begin{align}
    \mathbf{J}=-D_{\mathrm{eff}}(\mathbf{R})\nabla c=0\ \rightarrow\ c(\mathbf{R})=\mathrm{const}.
\end{align}

Next we consider an opposite situation, when the rates of spatial reactions and/or occupation probabilities vary in space, but diffusion constants $D^s$ do not depend on spatial coordinates. In this case, the drift velocity (\ref{velocity}) does not vanish. It can be moreover equivalently written as
\begin{align}\label{antidrift}
    \mathbf{V}=-\frac{\partial D_{\mathrm{eff}}}{\partial  \mathbf{R}}.
\end{align}
so that the evolution equation (\ref{anti2}) takes the form 
\begin{align}\label{gradient}
    \frac{\partial c}{\partial t}=\nabla^2\left[D_{\mathrm{eff}}(\mathbf{R})c \right].
\end{align}

According to this equation, in the steady state in absence of flux, we should have
\begin{align}\label{inverse}
    \mathbf{J}=-\nabla\left[D_{\mathrm{eff}}(\mathbf{R}) c\right]=0\ \rightarrow\ c(\mathbf{R})=\frac{A}{D_{\mathrm{eff}}(\mathbf{R})}
\end{align}
where $A$ is a normalization factor. Hence, molecules are depleted in the regions where their diffusion is fast.

As an example, we can again take the reaction cycle (\ref{cycle1}) with two states $s=0$ and $s=1$. At relatively small substrate concentrations, the turnover rate $k$ of this reaction and the occupation probability $\pi^1$ of the state $s=1$ are proportional to the substrate concentration $c_s$, i.e. we have $k=u c_s$ and $\pi^1=v c_s$ with some proportionality factors $u$ and $v$. If there is a spatial gradient of the substrate concentration, this leads to a gradient in the turnover rate, $\nabla k=u \nabla c_s$, and to a gradient in the occupation probability, $\nabla \pi^1=v \nabla c_s$.  Because the contributions to the diffusion coefficient are linear in $k$ and $\pi^1$, the drift velocity $\mathbf{V}=-\nabla D_\mathrm{eff}$ will be proportional to the substrate concentration gradient $\nabla c_s$ and directed oppositely to it. This can be described as an anti-chemotaxis effect.

Finally, if not only the reaction rates and occupation probabilities of different chemical states, but also the diffusion constants in such individual states depend on the particle position in space, the drift is present, but its velocity does not satisfy equation (\ref{antidrift}). Moreover, the evolution equation cannot be written in the simple anti-chemotaxis form (\ref{gradient}) and general evolution equation  (\ref{anti2}) must be always used in such general case.

\section{Numerical simulations \label{sec:simulation}}
Numerical simulations were performed for the model of deformable two-state dumbbells that was slightly different from that used in Section~\ref{sec:com_rc}. The stochastic dynamics of the dumbbell with two beads at positions $\textbf{R}_1$ and $\mathbf{R}_2$ was described by equations (\ref{LanvevinEq}). The interaction potential was chosen as
\begin{align}
U(r_{12}) = \frac{U_0}{2} \left( r_{12} - \ell_0 \right)^2
\end{align}
where $r_{12}=|\mathbf{R}_1-\mathbf{R}_2|$ and $\ell_0=2a$. 
Thus, explicitly, the dynamics equations were
\begin{align}
\gamma_1 \frac{d\mathbf{R}_1}{dt} = U_0 \left( \left| \mathbf{R}_2 - \mathbf{R}_1 \right| - \ell_0 \right)  \frac{\mathbf{R}_2 - \mathbf{R}_1}{\left| \mathbf{R}_2 - \mathbf{R}_1 \right|} + {\boldsymbol \xi}_1(t)
\end{align}
\begin{align}
\gamma_2 \frac{d\mathbf{R}_2}{dt} = U_0 \left( \left| \mathbf{R}_2 - \mathbf{R}_1 \right| - \ell_0 \right) \frac{\mathbf{R}_1 - \mathbf{R}_2}{\left| \mathbf{R}_1 - \mathbf{R}_2 \right|} + {\boldsymbol \xi}_2(t)
\end{align}
where
\begin{align}
\langle \xi_{i,\alpha}(t) \xi_{j,\beta}(s)\rangle = 2 \gamma_i k_BT \delta_{ij} \delta_{\alpha \beta} \delta(t-s)  
\end{align}
Two-dimensional systems were considered, i.e. $(\alpha,\beta)=(x,y)$.

In the simulation model, we have moreover dropped the assumption that transitions between the states $s=0$ and $s=1$ are possible only in a definite conformation (that is, at a certain distance $r_{12}$ between the beads). Instead, the transitions from $s=0$ to $s=1$ and back were allowed at rate constants $\kappa^{10}$ and $\kappa^{01}$ independent from the distance between the beads. We had $\gamma_1 = \gamma_2=\gamma$ in the state $s=0$ and $\gamma_1 = \gamma+\Delta \gamma$ and $\gamma_2 = \gamma-\Delta \gamma$ in the state $s=1$. Since the dumbbells were deformable and transitions could take place at any distance between the beads, there was a distribution of RC shifts and only the characteristic RC shift was given by equation (\ref{shift0}).

Note that the viscous friction coefficient $\Gamma=\gamma_1+\gamma_2$ was the same in the states $s=0$ and $s=1$. Therefore, the diffusion constants of the dumbbell were also the same in these states, i.e. $D^0=D^1$. Hence, if the enhancement of translational diffusion took place, it could only be due to the dancing effects. 

The parameters were fixed as $U_0 = 2$, $\ell_0 = 2a=1$, $\gamma = 1$, $k_BT = 1$, and $\kappa^{01}=10$; the remaining parameters $\Delta\gamma$ and $\kappa^{10}$ were varied in the simulations. Thus, the characteristic RC shifts were $d= \ell_0 \Delta \gamma / \Gamma = \Delta\gamma/2$.

The numerical integration was performed using the explicit Euler method with the fixed time step of $\Delta t = 10^{-3}$. The initial condition was set as $\mathbf{R}_1 = a\mathbf{e}_x$ and $\mathbf{R}_2 =- a \mathbf{e}_x$. Ensemble averages were calculated from the data for $10^7$ dumbbells. 

\begin{figure}
\centering
\includegraphics{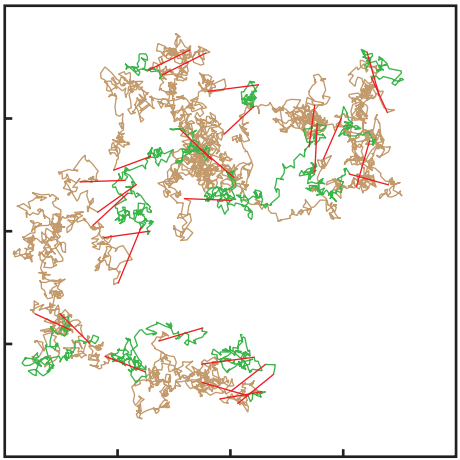}
\caption{The RC trajectory of the dumbbell over time $t=5$ within the window of length $\mathcal{L}=4$ (for $\Delta \gamma=0.8, \kappa^{10}=4,\kappa^{01}=10$, other parameters are given in the text). Fragments of the trajectory corresponding to the state $s=0$ are shown as green, those corresponding to $s=1$ are brown, and RC leaps in the transitions between the two states are displayed as red lines.}
\label{figNew}
\end{figure}

Figure~\ref{figNew} displays a typical RC trajectory of the dumbbell. Here, the parts of the trajectory corresponding to $s=0$ are shown in green and the parts corresponding to $s=1$ are in brown. The RC shifts in the (instantaneous) transitions between such two states are shown by red lines. Their lengths are different because the dumbbell in the simulation model was deformable, in contrast to the stiff dumbbell considered in Section~\ref{sec:com_rc}. 

\begin{figure}
\centering
\includegraphics{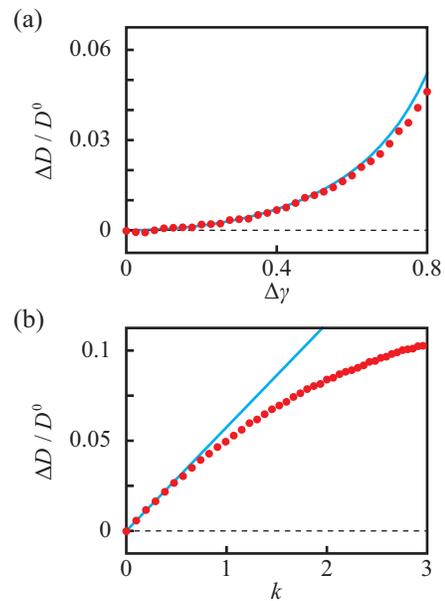}
\caption{Numerically determined relative diffusion enhancements $\Delta D/D^0= (D_\mathrm{eff}-D^0)/D^0$ as functions (a) of the parameter $\Delta \gamma$ (at $\kappa^{10}=1$) and (b) of the reaction turnover rate $k$ (at $\Delta\gamma=0.8$). To obtain the data in (b), simulations at $\kappa^{01}=10$ and for $\kappa^{10}$ varying from 0 to 4.2 were performed; the turnover rates $k$ were then determined by using equation (\ref{turnover}). For comparison, solid curves in (a,b) show the dependence predicted by equation (\ref{enhancement2}); numerical values for $D^1_{\mathrm{rot}}$ from Fig.~\ref{fig4} were here used.}
\label{fig3}
\end{figure}

The mean-square displacements (MSD) of CM were numerically determined. For all considered parameter values, MSD was linear in time. The effective diffusion coefficients could be calculated by dividing the MSD by $4t$ at $t=100$. Fig.~\ref{fig3}(a) shows the numerically obtained dependence of the relative diffusion enhancement $\Delta D/D^0$ on the model parameter $\Delta \gamma$ at $\kappa^{10}=1$.

\begin{figure}
\centering
\includegraphics{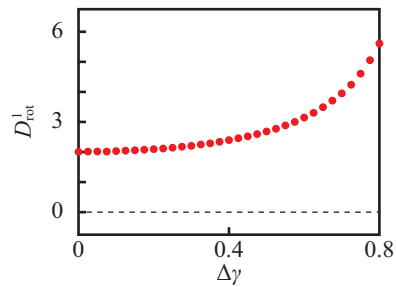}
\caption{Dependence of the rotational diffusion coefficient $D^1_\mathrm{rot}$ on the parameter $\Delta \gamma$ in the state $s=1$. For other parameters, see the text.}
\label{fig4}
\end{figure}

To interpret the simulation results, we additionally determined  rotational diffusion coefficients $D^s_\mathrm{rot}$ for the dumbbells. For the chosen parameter values, we found that $D^0_\mathrm{rot}=2.0$ in the state $s=0$. The rotational diffusion coefficient in the state $s=1$ increased with $\Delta\gamma$ as shown in Fig.~\ref{fig4}, reaching $D^1_\mathrm{rot}=5.6$ at $\Delta\gamma=0.8$. The orientational correlation time was   $\tau^0_\mathrm{rot}=1/D^0_\mathrm{rot}=0.5$  in the state $s=0$. Such time $\tau^1_\mathrm{rot}=1/D^1_\mathrm{rot}$ in the state $s=1$ decreased from 0.5 at $\Delta\gamma=0$ to 0.179 at $\Delta\gamma=0.8$.

Since, in the simulations, the characteristic lifetime of the state $s=1$ was equal to $\tau_1=1/\kappa^{01}=0.1$, this lifetime was comparable with the orientational correlation time in this state. In this case, the more general dependence, such as in equation (\ref{enhancement}), should be used:   
\begin{align}\label{enhancement2}
    D_{\mathrm{dance}}=\frac{1}{2}kd^2\frac{D^1_\mathrm{rot}}{\kappa^{01}+D^1_\mathrm{rot}}.
\end{align}
In contrast to equation (\ref{enhancement}), the prefactor is changed from 1/3 to 1/2 because the system is two-dimensional. Here, both $d$ and $D^1_\mathrm{rot}$ can depend on $\Delta\gamma$. Note that, in the considered model, we have $\Delta D = D_{\mathrm{dance}}$.

The theoretical estimate for the relative diffusion enhancement, based on the equation (\ref{enhancement2}) and using the dependence of $D^1_\mathrm{rot}$ on $\Delta \gamma$ from Fig.~\ref{fig4}, is shown by the solid curve in Fig.~\ref{fig3}a. We see that the agreement between the simulation data and the theoretical predictions is good. 

In Fig.~\ref{fig3}(b), the dependence of the diffusion enhancement on the reaction turnover rate $k$, based on the simulation data, is moreover displayed. In the considered model, the turnover rate is given by the equation
\begin{align}\label{turnover}
   k=\frac{\kappa^{10}\kappa^{01}}{\kappa^{01}+\kappa^{10}}. 
\end{align}
In the simulations used in Fig.~\ref{fig3}(b), the substrate binding rate $\kappa^{10}$ was varied between 0 and  4.2, whereas the rate constant of product formation was fixed at $\kappa^{01}=10$. Hence, the data for the range $0<k<3$ could be collected.

For comparison, the linear dependence, predicted by equation (\ref{enhancement2}), is shown by the solid line in Fig.~\ref{fig3}b. As we see, such linear dependence holds only at relatively small turnover rates. The difference at the higher turnover rates is due to correlations between the cycles. 

Indeed, in the derivation of equation (\ref{enhancement2}), it was assumed that dumbbell orientations in the next cycles were statistically independent, i.e. the memory of orientation was lost before the start of a new cycle. This implies that, for the linear dependence to hold,  the orientational correlation time $\tau^0_\mathrm{rot}$ in the state $s=0$ must be much shorter that the waiting time for substrate binding and, hence, initiation of a new cycle. Such waiting time $\tau_1=1/\kappa^{10}$ is inversely proportional to the substrate binding rate and gets shorter when it is increased. 

In the simulations in Fig.~\ref{fig3}(b), $\kappa^{10}$ was between 0 and 4.2, so that the shortest waiting time was $\tau_0=0.24$ at the largest considered turnover rate $k=3$. Hence, it was already comparable to the orientational correlation time $\tau^0_\mathrm{rot}=1/D^0_\mathrm{rot}=0.5$ in the state $s=0$. Therefore, the assumption of statistical independence of the cycles became violated at the higher turnover rates, leading to the deviations from the linear law (\ref{enhancement2}). 

Note that, because of equation (\ref{turnover}), the linear dependence of $\Delta D$ on $k$ implies the Michaelis-Menten dependence on the substrate binding rate. Therefore,  the deviations from the linear dependence mean that the Michaelis-Menten dependence of diffusion enhancement on the  substrate concentration holds only at relatively low concentrations of the substrate.

Additionally, we considered the situation when the reaction rate $\kappa^{10}$ depended on the position. We have set the size of the 2D system as $\mathcal{L}\times \mathcal{L}$ with $\mathcal{L} = 10$ and introduced the periodic boundary condition along both directions $x$ and $y$. We have chosen
\begin{align}
\kappa^{10} = \left\{ \begin{array}{ll} 2 & n\mathcal{L} - \Delta \mathcal{L} < y < n\mathcal{L} + \Delta \mathcal{L} \\ 0 & \textrm{otherwise}, \end{array} \right.
\end{align}
where $n$ is an integer. Hence, the reaction was taking place only within a central stripe of width $2\Delta \mathcal{L}$. 

To accumulate the data, simulations were performed for a system of $N=10^7$ independent dumbbells with the uniform initial distribution. The integration was carried out until $t = 100$ and the position distribution was obtained by averaging the distribution at every 0.1 time unit for $50 < t \leq 100$. 

\begin{figure}
\centering
\includegraphics{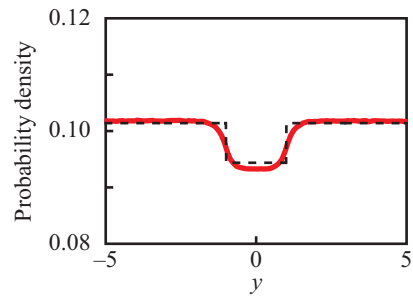}
\caption{Anti-chemotaxis. The final stationary particle distribution, projected on the $y$-axis, is shown. The substrate binding rate was set to be $\kappa^{10} = 2$ for $-\Delta\mathcal{L} < y < \Delta\mathcal{L}$ with $\Delta\mathcal{L}=2$, while it was zero outside of this interval; $\Delta\gamma=0.8$ and other parameters are the same as in Fig. 3. For comparison, the distribution $P(y)=C/D_\mathrm{eff}(y)$ is also displayed (dashed curve).}
\label{fig5}
\end{figure}

The obtained stationary distribution, projected on the $y$- axis, in the steady state  is shown in Fig.~\ref{fig5}. One can clearly see that the concentration is depleted in the central region where the reaction is present and, therefore, the diffusion is enhanced. Furthermore, the local concentration of dumbbells is approximately inversely proportional to the local diffusion coefficient, as predicted by equation (\ref{inverse}). This directly confirms the anti-chemotaxis effect,

It should be stressed that in the model used in numerical simulations, as well as in the general model in Section~\ref{sec:general}, rotational diffusion of the molecule results solely from translational diffusion of the beads (which have no own rotational friction). Thus, for a dumbbell, the rotational diffusion coefficient diverges in our model when the friction coefficient becomes vanishingly small for one of its beads (because it can then rotate without any friction around the second bead). In reality, the rotational diffusion constant of the dumbbell is bounded in this case by the remaining own rotational friction of the second bead.

However, the limit of diverging rotational diffusion has not been approached in our numerical simulations (as seen in Fig.~\ref{fig4}, the rotational diffusion coefficient of the dumbbell could only increase by about a factor of 2 within the considered parameter range). Moreover, as shown in Section~\ref{sec:dancing}, diffusion boosting ceases to depend on the rotational diffusion constant of the dumbbell, provided that it is large enough (see equation~\eqref{enhancement}); therefore, the correct value of such constant is not important in this case. Note furthermore that, in the related research on systems of self-propelled dumbbells, the own rotational viscous friction of the beads is often neglected too \cite{suma,winkler,kumar}.

\section{Discussion: Boosted diffusion of enzymes \label{sec:discussion}}
In several enzymes, such as urease, catalase or acetylcholinesterase (AChE), substantial diffusion enhancement by tens of percent has been reported under reaction conditions by diverse research groups \cite{dey-review,bustamante, granick}.  The common feature of these enzymes is that they have high turnover rates of about $k=10^5\,\mathrm{s}^{-1}$.  Moreover, they are highly exergonic, with the net Gibbs free energy release $\Delta G$ by tens of $k_BT$ within a catalytic cycle\cite{granick-master}. They are also oligomeric: urease forms rings with six catalytically active monomers, catalase is a tetramer and AChE is a dimer. Weak diffusion enhancement by a few percent has been seen for some enzymes with moderate Gibbs free energy release and typical (slow) turnover rates too\cite{granick-master,granick-science}.  According to the detailed study \cite{granick-master}, boosted diffusion is absent for endergonic reactions, i.e. for the reactions that are not able to generate mechanical work. All known reactions with boosted diffusion are exothermic, i.e. releasing heat.  Diffusion enhancement has been  once reported \cite{exo} for a slow endothermic reaction catalyzed by aldolase, but this observation could not be later confirmed \cite{hess}.

Super-resolution optical microscopy experiments have suggested that, under highly boosted diffusion, the enzymes perform sudden leaps by about tens of nanometers within tens of microseconds and such leaps are superimposed on the classical thermal Brownian motion \cite{granick}. The leaps, that could not be directly observed, but were deduced by the statistical analysis of experimental data, had lengths of $50$--$100\,\textrm{nm}$. Note however that, as mentioned by the authors\cite{granick}, these values represented ``upper bounds of leap length because what we express here as ballistic speed likely consists of a sequence of shorter steps that have not yet randomized.'' Hence, elementary leaps (``steps'') could have been actually shorter, i.e. of the order of the typical protein molecule size of $10\,\textrm{nm}$. Remarkably, the intervals between the leaps were of about the same duration as the turnover times.

Examination of further data for various enzymes has revealed that diffusion enhancement is linearly proportional to the turnover rate \cite{granick-master}. Moreover, a linear dependence of diffusion enhancement on the Gibbs energy release per a turnover cycle holds \cite{granick-master}. When a spatial gradient of the substrate concentration is present, the enzymes drift in the direction opposite to it, which has been described as an anti-chemotaxis effect \cite{granick}. Furthermore, the local enzyme concentration in the steady state is inversely proportional to the local diffusion coefficient \cite{granick}.

To interpret such observations, it has been repeatedly suggested that some form of molecular self-propulsion is involved \cite{dey-review,granick}. However, no feasible mechanism of sufficiently strong self-propulsion could be identified so far.

Furthermore, there is a common problem under \textit{any} self-propulsion mechanism. The velocity of a self-propelling particle in the low Reynolds number regime should be proportional to the rate at which energy is supplied to it. Hence, leap lengths should be also proportional to this rate and the mean-square displacement, determining the diffusion constant, must be then quadratic in the rate of energy release. But, as already noted, the actual diffusion boosting is a linear function of the energy release rate \cite{granick-master}. 

It has been also noted \cite{agudo,mandal, agudoC} that boosted diffusion can arise because more compact, and thus more mobile, enzyme conformations might become occupied under a reaction, leading, in the case of two-state models, to an increase of the diffusion coefficient corresponding to the first term in equation (\ref{two}). This has already allowed to reproduce the anti-chemotaxis effect, but has left not explained the characteristic molecular leaps \cite {granick} and the reported strong difference in boosting between exergonic and endergonic enzymes\cite{granick-master}.

Reaction-induced molecular dancing, considered by us, could be instead responsible for the observed boosted diffusion of enzymes. Indeed, principal aspects of the observed effects become thus reproduced: 

\begin{itemize}
    \item Under molecular dancing, an enzyme performs the leaps of a length comparable to its size in each turnover cycle. For oligomeric structures, the size of a leap can be comparable to structure's length.
    \item Such independent leaps in random directions lead to diffusion enhancement proportional to the turnover rate (note that, even for fast enzymes, turnover times are still longer than orientational correlation times),
    \item The anti-chemotaxis effect is predicted, with the drift of enzymes against the substrate concentration gradient. The local enzyme concentration in the steady state is inversely proportional to the local diffusion coefficient.
    \item Numerical estimates can be made by using the dependence $\Delta D=(1/3)kd^2$ for the model dumbbell enzymes.  If the RC shift $d$ within the reaction cycle is comparable to the enzyme size, $d=10$ nm can be taken. Then, for enzymes with high turnover rates $k=10^5$ $\mathrm{s}^{-1}$, the diffusion enhancement about $\Delta D=10^{-7}$ $\mathrm{cm}^2/\mathrm{s}$ is obtained,  which is comparable to the typical diffusion constants of proteins and is consistent with the reported maximum diffusion enhancement by tens of percent \cite{dey-review,granick}.
    \item Typical rotation diffusion constants for proteins, as yielded both by MD simulations and NMR measurements, are of the order of $D_\mathrm{rot}=10^7\,\mathrm{s}^{-1}$ (see ref.\cite{iwai}). Hence, even for the fastest enzymes with the turnover rates about $k=10^5\,\mathrm{s}^{-1}$, the condition $D_\mathrm{rot}\gg k$ would be satisfied, i.e. the enzyme direction would be randomized between the cycles. For mesoscopic enzymic structures, such as the urease rings, rotational diffusion can be however slower, so that subsequent leaps (``steps'') are not yet completely randomized. The regimes without complete randomization of dumbbell orientations were also observed in our numerical simulations in Section~\ref{sec:simulation} when reaction rates $k$ were increased.
\end{itemize}

There are also several issues that need to be further discussed:
\begin{itemize}
\item The Protein Data Base (PDB) does not show significant ligand-induced conformational changes in the enzymes where substantial boosting takes place. Hence, there is no structural evidence for the existence of a state with a largely different enzyme shape, as would have been necessary for the dancing effects. 

However, the PDB data is based on X-ray diffraction experiments with crystallized proteins and therefore it can contain only stable states (in the case of a reaction, they are typically yielded by employing non-reactive ligand analogs). Such data does not include information on the transient states. 

As already noted, enzymes with substantial diffusion enhancement tend to be oligomeric. Urease, where the strongest boosting was seen, is a hexamer with the rings formed by six catalytically active identical subunits. The amount of energy released in each subunit within a turnover cycle in this enzyme is so high that even reaction-induced breakup of the rings has been suggested \cite{zhang}. Although the breakup has been experimentally excluded \cite{xu}, transient deformations of the rings could still be strong. For oligomeric enzymes, the transient state with a short lifetime can correspond to the excitation of a normal mode representing a deformation of the oligomeric superstructure, such as the urease ring, immediately after the catalytic reaction event.

To roughly account for this in the considered minimal description, a modified model of an \textit{excitable} dumbbell with the reaction scheme
\begin{align}
    S+E\rightarrow EP^*\rightarrow EP \rightarrow E+P.
\end{align}
can be introduced. This dumbbell has a transient excited state $EP^*$. One can assume that only in this transient state of the enzyme-product complex, the friction coefficients of the beads are changed. For example, we can choose $\gamma_1=\gamma+\Delta \gamma$ and $\gamma_2=\gamma-\Delta \gamma$ in the state $EP^*$ and equal to $\gamma$ in all other states. 

Diffusion enhancement for such excitable dumbbell is determined by the general estimate (\ref{deltaD}). If the transient excitable state has a short life-time, its occupation probability is low, so that the first term in equation (\ref{deltaD}) is small. Moreover, only the transitions $E+S\rightarrow EP^*$ and $EP^*\rightarrow EP$ will contribute to the second term. Hence, the diffusion enhancement is still given by $\Delta D=(1/3)kd^2$ where $k$ is the turnover rate and $d$ is the shift of the rotation center in these two transitions. Therefore, the above numerical estimates continue to be applicable in this case, even in absence of a (meta)stable ligand-bound enzyme state, 

\item Moreover, the reported diffusion boosting was proportional to the difference $\Delta G$ in Gibbs free energies between the substrate and the product, but such dependence appears to be absent in equation (\ref{eff}) for diffusion enhancement due to molecular dancing. 

The dependence enters, however, into the theory when elastic deformations are explicitly introduced. Generally, the elastic deformation energy contained in a normal mode with amplitude $A$ is $\Delta \mathcal{E}=(1/2)KA^2$ where $K$ is the mode stiffness constant. Elastic deformations in the enzyme lead to changes in friction coefficients of its domains and therefore to shifts in the rotation center. It can be expected that the magnitude of an RC shift would be proportional to the amplitude of the excited normal mode, i.e. that $d\propto A$. 

This implies that, for the change in the diffusion coefficient, we would have $\Delta D \propto A^2\propto \Delta \mathcal{E}$. Hence, if the elastic deformation energy $\Delta \mathcal{E}$ represents a dominant contribution into the change $|\Delta G|$ of the Gibbs free energy under the reaction, diffusion enhancement should be linear in $\Delta G$, as indeed observed \cite{granick-master}. Note furthermore that, if the excitation energy is stored in elastic deformations, it can be further used to perform work. This means that the respective chemical reactions would be exergonic, also in agreement with the experimental results \cite{granick-master}.

For oligomeric enzymes, large elastic deformations of their mesoscopic multi-unit structures (such as the urease ring) can follow after a catalytic event with much energy release in one of the subunits. For example, the excited normal mode in urease can correspond to a strong deformation of the ring. Then, its amplitude $A$ and the RC shift $d$, determining the leap length, would be of the same mesoscopic order of magnitude as the diameter of the ring.

\end{itemize}

While the arguments presented in this Section make very likely an explanation of boosted diffusion in terms of molecular dancing effects,   quantitative treatment of the proposed effects for specific enzymes and  detailed comparison of the predictions to the respective experimental results remain to be undertaken. This should be a subject of future work. Additional experiments to test the theory need to be performed.

\section{Conclusions \label{sec:conclusion}}

In this study, a novel mechanism of reaction-induced motion for enzymes has been proposed and analyzed. In contrast to previously known mechanisms, it does not involve self-propulsion. Instead, it is based on the effect of molecular dancing introduced by us. Both a general theory of these phenomena has been constructed and the analysis for a simple model of the two-state dumbbell has been performed. Theoretical predictions have been checked in direct numerical simulations.

As we have found, a statistical description in terms of molecular centers of mass does not provide a good basis for considering the reaction effects. This is because, in solutions, the center of mass is not conserved even in absence of the reactions and hydrodynamic interactions, i.e. only under potential interactions between the domains. In contrast to this, separation of variables in absence of the reactions and hydrodynamic interactions takes place in the description in terms of the rotation center of a molecule. 

Hence, the position of the rotation center (but not of the center of mass) represents a \textit{slow} variable similar, e.g.,  to the oscillation phase. When chemical reactions are added, this variable undergoes, in presence of rotational diffusion, an additional translational random walk. Remarkably, while the two systems are physically much different, the mathematical theory of molecular dancing resembles the classical Kuramoto theory \cite{kuramoto} for phase drift in populations of coupled oscillators. 

It has been previously known that reactions generate additional noise, but only the enhancement of concentration fluctuations due to the internal noise of reactions has been considered so far (see, e.g., ref.\cite{gardiner, mikhailovI}). In the present study, we have essentially demonstrated that coupling between the mass center and changes in molecular conformations generates a different kind of reaction noise that enhances spatial diffusion of molecules. Note that, similar to the previously known effects of internal reaction noise\cite{gardiner, mikhailovI}, both equilibrium and nonequilibrium reactions can boost molecular diffusion. However, it can be expected that, for equilibrium reactions, this would be a weak effect.

Our analysis strongly suggests that molecular dancing has been responsible for boosted diffusion observed for various enzymes. Additional investigations and experimental tests are however needed to confirm this. 

In the present study, hydrodynamic effects for proteins were taken into account in the simplest Langevin description, with the viscous friction coefficient dependent on the shape of a protein domain. Hydrodynamic interactions between the domains (beads) are not principal for the considered dancing mechanism and they have been neglected by us. For completeness, they should still be, however, taken into account in future theoretical work.

The attention in this study was focused on enzymes. However, the constructed theory is more general and it can be applied to other systems, natural or artificial, as well. Synthetic micro-dancers can be designed and used to test directly the predictions of boosted diffusion. Various biotechnology applications with synthetic micro-dancers can be developed too.

\begin{acknowledgments}

The authors thank R.~Kapral and Y.~Koyano for interesting discussions. This study was supported by the Japanese Society for Promotion of Science through Grants-in-Aid for Scientific Research (C) JP19K03765  (A.M. and H.K.) and (B) JP21H01004 (H.K.). A.M. wishes to use this opportunity to express his deep gratitude to the staff and colleagues for support during his work at the World Premier Initiative Nano Life Science Institute in Kanazawa.

\end{acknowledgments}

\appendix
\section{Derivation details of the general theory}

In this Appendix, a detailed formulation of the theory of reaction-induced molecular dancing is presented. 

\subsection{Stochastic Langevin dynamics of CM and RC in absence of reactions}

In absence of reactions, the master equation (\ref{master}) corresponds to a set of stochastic Langevin equations
\begin{align}
\gamma^s_i\frac{d R_{i,\alpha}}{dt}=-\sum_j\frac{\partial U^s_{ij}}{\partial R_{i,\alpha}}+\xi_{i,\alpha}(t)
\end{align}
with independent thermal noises, such that
\begin{align}
    \langle \xi_{i,\alpha}(t)\xi_{j,\beta}(t')\rangle=2\gamma^s_ik_BT\delta_{ij}\delta_{\alpha \beta}\delta(t-t').
\end{align}

The CM position of a molecule is $\mathbf{R}=(1/N)\sum_{i}\mathbf{R}_i$ and the deviations from it are $\mathbf{r}_i=\mathbf{R}_i-\mathbf{R}$. Stochastic Langevin equations for these variables are
\begin{align}\label{LangevinCM}
\frac{d R_{\alpha}}{dt}=-\frac{1}{2N}\sum_{i,j}(\mu^s_i-\mu^s_j)\frac{\partial U^s_{ij}}{\partial r_{i,\alpha}}+\chi_{\alpha}(t)
\end{align}
\begin{align}
\frac{d r_{i,\alpha}}{dt}=-\sum_{j}\mu^s_i\frac{\partial U^s_{ij}}{\partial r_{i,\alpha}}+\frac{1}{2N}\sum_{i,j}(\mu^s_i-\mu^s_j)\frac{\partial U^s_{ij}}{\partial r_{i,\alpha}}+\rho_{i,\alpha}(t)
\end{align}
with thermal noises $\chi_\alpha(t)=(1/N)\sum_i\mu^s_i\xi_{i,\alpha}(t)$ and $\rho_{i,\alpha}(t)=\mu^s_i\xi_{i,\alpha}(t)-\chi_\alpha(t)$. The correlation functions of the noises are 
\begin{align}
    \langle \chi_{\alpha}(t)\chi_\beta(t')\rangle=\frac{2}{N}\bar{\mu}^sk_BT\delta_{\alpha \beta}\delta(t-t').
\end{align}
\begin{align}\label{crossCM}
    \langle \chi_{\alpha}(t)\rho_{i,\beta}(t')\rangle=\frac{2}{N}(\mu^s_i-\bar{\mu}^s)k_BT\delta_{\alpha \beta}\delta(t-t').
\end{align}
\begin{align}\label{rhoCorr}
    \langle \rho_{i,\alpha}(t)\rho_{j,\beta}(t')\rangle=&2k_BT\Big(\mu^s_i\delta_{ij}-\frac{1}{N}\mu^s_i-\frac{1}{N}\mu^s_j+\frac{1}{N}\bar{\mu}^s\Big) \nonumber \\
    & \times \delta_{\alpha \beta}\delta(t-t')
\end{align}
where $\bar{\mu}^s=(1/N)\sum_i\mu^s_i$ is the mean mobility of the beads in the chemical state $s$.

Unless the mobilities of all beads are equal, the motion of CM is thus affected by conformational intramolecular dynamics and by rotations of the molecule. The intramolecular forces enter into the motion equation (\ref{LangevinCM}) for CM. Moreover, as seen in equation (\ref{crossCM}), the CM noise $\boldsymbol{\chi}(t)$ is correlated with the intramolecular noises $\boldsymbol{\rho}_i(t)$.

The RC position of a molecule in the state $s$ is  $\mathbf{Q}^s=\sum_{i}(\gamma^s_i/\Gamma^s)\mathbf{R}_i$, where $\Gamma^s=\sum_i\gamma^s_i$, and the deviations from it are $\mathbf{q}^s_i=\mathbf{R}_i-\mathbf{Q}^s$. 

It can be verified that $\mathbf{Q}^s(t)$ indeed represents the position of the center around which rotations take place. To induce rigid translational motion of the entire molecule at velocity $\mathbf{V}$, forces $\mathbf{F}^s_i=-\gamma^s_i\mathbf{V}$ need to be applied to individual beads. The total torque $\mathbf{N}^s$, created by such forces with respect to the position $\mathbf{Q}^s$, is zero, as it should be for the RC. We have  $\mathbf{N}^s=\sum_i\big[\mathbf{F}^s_i\times (\mathbf{R}_i-\mathbf{Q}^s)\big]=-\sum_i\gamma^s_i\big[\mathbf{V}\times (\mathbf{R}_i-\mathbf{Q}^s)\big]=0.$

Stochastic Langevin equations for the RC  variables are
\begin{align}\label{LangevinRC}
\Gamma^s\frac{d Q^s_{\alpha}}{dt}=\eta_{\alpha}(t),
\end{align}
\begin{align}\label{Langevinq}
\gamma^s_i\frac{d q^s_{i,\alpha}}{dt}=-\sum_{j}\frac{\partial U^s_{ij}}{\partial q^s_{i,\alpha}}+\zeta_{i,\alpha}(t)
\end{align}
with thermal noises $\eta_\alpha(t)=\sum_i\xi_{i,\alpha}(t)$ and $\zeta_{i,\alpha}(t)=\xi_{i,\alpha}(t)-(\gamma^s_i/\Gamma^s)\eta_\alpha(t)$. The correlation functions of the noises are 
\begin{align}
    \langle \eta_{\alpha}(t)\eta_\beta(t')\rangle=2\Gamma^sk_BT\delta_{\alpha \beta}\delta(t-t'),
\end{align}
\begin{align}\label{crossRC}
    \langle \eta_{\alpha}(t)\zeta_{i,\beta}(t')\rangle=0,
\end{align}
\begin{align}\label{zetaCorr}
    \langle \zeta_{i,\alpha}(t)\zeta_{j,\beta}(t')\rangle=2\gamma^s_i\Big(\delta_{ij}-\frac{\gamma^s_j}{\Gamma^s}\Big)k_BT\delta_{\alpha\beta}\delta(t-t').
\end{align}

Remarkably, the motion of RC is decoupled from the intermolecar dynamics and it is also not affected by rotations of a molecule. The intramolecular forces do not enter into the equation of motion (\ref{LangevinRC}) for RC. Moreover, as seen in equation (\ref{crossRC}), the RC noise $\eta(t)$ is independent from the intramolecular noises $\zeta_i(t)$. Thus, thermal rotational fluctuations take place around RC.

According to equation (\ref{zetaCorr}), intramolecular noises acting on different beads (when $i\neq j$) are anti-correlated. Such cross-correlations arise because $N$ stochastic variables $\mathbf{q}^s_i(t)$ must satisfy the constraint 
\begin{align}\label{constraint}
 \sum_i\frac{\gamma^s_i}{\Gamma^s}\mathbf{q}^s_i=0.   
\end{align}

\subsection{Master equation for RC}

The system is characterized by a set of $(N+1)$-dimensional probability distributions $\{p^s_{RC}(\mathbf{Q}^s,\{\mathbf{q}^s_k\},t)\}$. The RC distributions in different chemical states are obtained by averaging, i.e. as
\begin{align}\label{RCdist}
    P_{RC}^s(\mathbf{Q}^s,t)=\int p_{RC}^s(\mathbf{Q}^s,\{\mathbf{q}^s_k\},t)d^N\mathbf{q}^s_k.
\end{align}

In absence of reactions, the master equation follows from the Langevin equations (\ref{LangevinRC})-(\ref{zetaCorr}). If a Langevin equation for a stochastic variable is known, the Fokker-Planck equation for the probability distribution of this variable can be constructed (see ref.\cite{risken}). In the case of equations (\ref{LangevinRC})-(\ref{zetaCorr}), such general construction yields
\begin{align} \label{rot}
    \frac{\partial p^s_{RC}}{\partial t}=-\frac{\partial H^s_{\alpha}}{\partial Q^s_{\alpha}}-\sum_i\frac{\partial h^s_{i,\alpha}}{\partial q^s_{i,\alpha}}
\end{align}
where
\begin{align}\label{driftH}
    H^s_\alpha=-k_BT M^s\frac{\partial p^s_{RC}}{\partial Q^s_\alpha},
\end{align}
and 
\begin{align} \label{eqnA}
    h^s_{i,\alpha}=& -\mu^s_i\left(\sum_j\frac{\partial U^s_{ij}}{\partial q^s_{i,\alpha}}p^s_{RC}+k_BT\frac{\partial p^s_{RC}}{\partial q^s_{i,\alpha}}\right) \nonumber \\
    & +k_BTM^s\sum_j\frac{\partial p^s_{RC}}{\partial q_{j,\alpha}}.
\end{align}
Here, $M^s$ is the mobility of the whole molecule,
\begin{align}\label{mobil}
    M^s=\frac{1}{\Gamma^s}=\frac{1}{\sum_i(1/\mu^s_i)}.
\end{align}
The last term in equation (\ref{eqnA}) is due to the constraint (\ref{constraint}).

Next, reaction terms can be added into the master equation (\ref{rot}). In doing this, it should be taken into account that the RC is different in different chemical states. The shift of RC in a transition from $s'$ to $s$ is
\begin{align}\label{change}
\mathbf{d}^{ss'}=\mathbf{Q}^s-\mathbf{Q}^{s'}=\sum_i\Big(\frac{\gamma^s_i}{\Gamma^s}-\frac{\gamma^{s'}_i}{\Gamma^{s'}}\Big)\mathbf{r}_i=-\frac{1}{\Gamma^{s}}\sum_{i'} \gamma^s_{i'} \mathbf{q}^{s'}_{i'}.    
\end{align}
Thus, a reaction event $s'\rightarrow s$ is accompanied by a shift in the coordinates $\mathbf{Q}^{s'}\rightarrow \mathbf{Q}^s$, $\mathbf{q}^{s'}_k\rightarrow\mathbf{q}^s_k$, such that 
\begin{align}
    \mathbf{Q}^{s'} =\mathbf{Q}^s-\mathbf{d}^{ss'},\ \mathbf{q}^{s'}_k =\mathbf{q}^s_k+\mathbf{d}^{ss'}.
\end{align}

When the reactions are included, the master equation is
 \begin{align}
& \frac{\partial p^s_{RC}}{\partial t}=-\frac{\partial H^s_{\alpha}}{\partial Q^s_{\alpha}}-\sum_i\frac{\partial h^s_{i,\alpha}}{\partial q^s_{i,\alpha}} \nonumber \\
&+ \sum_{s'}\left[\int  w^{ss'}\prod_{i,j}\delta\left(\left|\mathbf{q}^{s'}_i-\mathbf{q}^{s'}_j\right|-r_{ij}^{ss'}\right) \right. \nonumber \\
& \quad \left. \times \delta\left(\mathbf{Q}^s-\mathbf{Q}^{s'}+\frac{1}{\Gamma^s}\sum_{i'}\gamma^s_{i'}\mathbf{q}_{i'}^{s'}\right) \right. \nonumber \\
& \quad \left. \times \prod_k\delta\left(\mathbf{q}^s_k-\mathbf{q}_k^{s'} -\frac{1}{\Gamma^s}\sum_{i'}\gamma^s_{i'}\mathbf{q}_{i'}^{s'}\right) \right. \nonumber \\ 
& \quad \left. \times p^{s'}_{RC}(\mathbf{Q}^{s'},\{\mathbf{q}^{s'}_k\})d\mathbf{Q}^{s'}d^N\mathbf{q}^{s'}_k \right. \nonumber \\ &
\left. -w^{s's} \prod_{i,j}\delta\left(\left|\mathbf{q}_i^s-\mathbf{q}^s_j\right|-r_{ij}^{s's}\right)p^s_{RC}\left(\mathbf{Q}^s,\{\mathbf{q}^s_k\}\right)\right]
 \end{align}
 
Integration over $\mathbf{Q}^{s'}$ can be performed and, after that, we obtain the evolution equations in the final simple form:
 \begin{align}\label{master-rot}
& \frac{\partial p^s_{RC}}{\partial t}=-\frac{\partial H^s_{\alpha}}{\partial Q^s_{\alpha}}-\sum_i\frac{\partial h^s_{i,\alpha}}{\partial q^s_{i,\alpha}} \nonumber \\ 
& + \sum_{s'}\left[\int  w^{ss'}\prod_{i,j}\delta\left(\left|\mathbf{q}^{s'}_i-\mathbf{q}^{s'}_j \right|-r_{ij}^{ss'}\right) \right. \nonumber \\
& \quad \times \left. \prod_k\delta\left(\mathbf{q}^s_k-\mathbf{q}_k^{s'}+\mathbf{d}^{ss'}\right) \right. \nonumber \\
&\quad \left. \times p^{s'}_{RC}\left(\mathbf{Q}^s-\mathbf{d}^{ss'},\{\mathbf{q}^{s'}_k\}\right)d^N\mathbf{q}^{s'}_k \right. \nonumber \\
 & \left. -w^{s's}\prod_{i,j}\delta\left(\left|\mathbf{q}^s_i-\mathbf{q}^s_j\right|-r_{ij}^{s's}\right)p^s_{RC}\left(\mathbf{Q}^s,\{\mathbf{q}^s_k\}\right)\right]
 \end{align}
 where $\mathbf{d}^{ss'}$ is given by equation (\ref{change}).
 
 This result has a simple interpretation. The first term on the right side describes free diffusion motion of RC. The second term accounts for rotational diffusion and conformational fluctuations. The last term corresponds to chemical reactions. A reaction from the state $s'$ to the state $s$ and back is allowed only in a definite conformation specified by a set of pair distances $r^{ss'}_{ij}$ between the beads. The molecule in such conformation can, however, be still arbitrarily rotated in space. Each chemical reaction $s'\rightarrow s$ involves a shift in the RC position by $\mathbf{d}^{ss'}$ that depends on the conformation and orientation of the molecule before the transition event.

 \subsection{Systems without reactions}
 When the reaction terms are absent, the master equation (\ref{master-rot}) becomes 
 \begin{align} \label{NoReaction}
    \frac{\partial p^s_{RC}}{\partial t}=-\frac{\partial H^s_{\alpha}}{\partial Q^s_{\alpha}}-\sum_i\frac{\partial h^s_{i,\alpha}}{\partial q^s_{i,\alpha}}.
\end{align} 
 Note that, without the reactions, the molecule is always in the same chemical state.  
 
 In this evolution equation, separation of variables $\mathbf{Q}^s$ and $\mathbf{q}^s_i,i=1,\dots ,N$, takes place. The RC probability flux $\mathbf{H}$ depends only on $\mathbf{Q}^s$, whereas the fluxes $\mathbf{h}^s_i$ are independent from the RC position. By averaging over the set of $\mathbf{q}^s_i$ variables, a closed evolution equation for RC is derived:
 \begin{align} \label{RCdiff}
    \frac{\partial P^s_{RC}}{\partial t}=D^s\frac{\partial^2 P^s_{RC}}{\partial \left(Q_{\alpha}^s\right)^2}
\end{align} 
  where the diffusion constant is $D^s=k_BTM^s$ and $M^s=1/\big(\sum_i\gamma^s_i\big)$ is the mobility coefficient of the entire molecule consisting of $N$ beads. Remarkably, $M^s$ is the same as the mobility of a \textit{rigid} molecule with the beads of individual mobility $\mu^s_i$. Note that this is an exact evolution equation, because no approximations were made to derive it from the master equation (\ref{NoReaction}) for the joint probability distribution. 
  
As follows from our analysis based on the Langevin equations, there is no separation of the CM variables $\mathbf{R}$ and $\mathbf{r}_i,i=1,\dots ,N$. Therefore, an \textit{exact} closed evolution equation for the CM probability distribution $P_{CM}(\mathbf{R})$, averaged over $\mathbf{r}_i$, cannot be derived even in absence of chemical reactions. Indeed, on the short timescales, characteristic for conformational and orientational fluctuations, dancing of CM takes places. 
 
 Nonetheless, if we consider a smooth CM probability distribution $\tilde{P}_{CM}$, averaged over time intervals longer than such short timescales, its evolution will be given by equation
 \begin{align} \label{CMdiff}
    \frac{\partial P^s_{CM}}{\partial t}=D^s\frac{\partial^2 P^s_{CM}}{\partial R_{\alpha}^2}
\end{align}
 with the same diffusion coefficient $D^s$ as in equation (\ref{RCdiff}). 
 
 This follows from the fact that the two collective variables $\mathbf{R}(t)$ and $\mathbf{Q}^s(t)$ differ only by a term of the order of the size of a molecule and the magnitude of the difference cannot indefinitely increase with time. Hence, the long-time limits $\lim_{t\rightarrow \infty} (1/t)\langle \big(\Delta \mathbf{R}(t)\big)^2\rangle$ and $\lim_{t\rightarrow \infty }(1/t)\langle \big(\Delta \mathbf{Q}^s(t)\big)^2\rangle$ for their mean-square displacements must coincide, implying that the diffusion constants are the same.
 
For thermal equilibrium at temperature $T$ in absence of reactions, the system is described by the Boltzmann probability distribution 
 \begin{align}\label{Boltzmann}
     \rho^s_{RC}(\{\mathbf{q}_i\})=\frac{1}{Z^s}\exp \left(-\frac{1}{2k_BT}\sum_{i,j}U^s_{ij}(\mathbf{q}^s_i-\mathbf{q}^s_j)\right)
 \end{align}
where $Z^s$ is the normalization factor. It can be checked that this distribution is a solution of the master equation (\ref{rot}). Note that the contribution from the last term in the flux $\mathbf{h}^s_i$ vanishes because, for the Boltzmann distribution, we have 
\begin{align}
    \sum_{i,j}\frac{\partial^2 \rho^s_{RC}}{\partial q^s_i\partial q^s_j}=0.
\end{align}

The equilibrium distribution (\ref{Boltzmann}) depends only on distances between the beads. Such distances determine a conformation, but leave arbitrary the orientation of the molecule. The  equilibrium state is statistically isotropic, with all molecular orientations having the same statistical weight.

\subsection{Approximate evolution equation for systems with reactions}

Above, exact evolution equations (\ref{master-rot}) for joint RC probability distribution for systems with reactions has been derived. Now, we want to obtain from it an approximate evolution for reduced RC probability distributions (\ref{RCdist}) .

By averaging equations (\ref{master-rot}) over  variables $\mathbf{q}^s_i$, the following system of equations for reduced distributions $P^s_{RC}$ is found:
\begin{align}
& \frac{\partial P^s_{RC}}{\partial t}=D^s\frac{\partial^2 P^s_{RC}}{(\partial Q^s_\alpha)^2} \nonumber \\
&+ \sum_{s'}\left[\int w^{ss'}\prod_{i,j}\delta\left(\left|\mathbf{q}^{s'}_i-\mathbf{q}^{s'}_j\right|-r_{ij}^{ss'}\right) \right. \nonumber \\ & \left. \quad \times p^{s'}_{RC}\left(\mathbf{Q}^s+\frac{1}{\Gamma^s}\sum_k\gamma^s_k\mathbf{q}_k^{s'},\{\mathbf{q}^{s'}_k\}\right)d^N\mathbf{q}_k^{s'} \right. \nonumber \\
& \left. -\int w^{s's}\prod_{i,j}\delta\left(\left|\mathbf{q}^s_i-\mathbf{q}^s_j\right|-r_{ij}^{s's}\right)p^s_{RC}\left(\mathbf{Q}^s,\{\mathbf{q}^s_k\}\right)d^N\mathbf{q}^s_k\right].
 \end{align}

This system of equations for $P^s_{RC}(\mathbf{Q}^s,t)$ is not closed:  reaction terms include averages taken with joint probability distributions $p^s_{RC}(\mathbf{Q}^s,\{\mathbf{q}^s_k\})$. However, under certain conditions, approximate closed evolution equations can be derived.

The considered system possesses several distinct relaxation timescales. The characteristic timescale $\tau_c$ specifies the time within which conformational equilibrium within a molecule is reached. The orientational correlation time $\tau_\mathrm{{rot}}$ is the time within which the memory of initial orientation is lost. The timescale $\tau_r$ is the characteristic time within which  chemical equilibrium is achieved.

In our approximate derivation, we shall assume that the conditions $\tau_r\gg\tau_{\mathrm{rot}}$ and $\tau_r \gg \tau_c$ are always satisfied. Hence, orientational and conformational equilibration represent the fastest processes. Reactions produce correlations between the instantaneous RC position and conformational/orientational molecular states. Such correlations however persist only over a very short time. Because rotational diffusion is fast, the molecule randomly tumbles between any two reaction events, so that the memory of the initial orientation becomes lost.

Under such conditions, local conformational equilibrium is maintained. Therefore, probability distributions, entering into the reaction terms, can be approximately factorized
\begin{align}\label{factor}
   p^s_{RC}(\mathbf{Q}^s,\{\mathbf{q}^s_k\},t)= P^s_{RC}(\mathbf{Q}^s,t)\rho^s_{RC}(\{\mathbf{q}^s_k\})
\end{align}
where $\rho^s_{RC}$ is given by equation (\ref{Boltzmann}).
 Note that
\begin{align}
    \sum_s \int P^s_{RC}(\mathbf{Q}^s)d\mathbf{Q}^s=1.
\end{align}

 Substituting this and integrating over variables $\mathbf{q}^s_k$, a closed set of evolution equations for RC distributions becomes obtained:
\begin{align}\label{master-closed}
  & \frac{\partial P^s_{RC}}{\partial t}=D^s\frac{\partial^2 P^s_{RC}}{(\partial Q^s_\alpha)^2} \nonumber \\
  & + \sum_{s'}\left[ \int w^{ss'}\prod_{i,j}\delta\left(\left|\mathbf{q}^{s'}_i-\mathbf{q}^{s'}_j\right|-r_{ij}^{ss'}\right) \right. \nonumber \\ & \quad \times \left. P^{s'}_{RC}\left(\mathbf{Q}^s-\mathbf{d}^{ss'}\right)\rho^{s'}_{RC}\left(\{\mathbf{q}^{s'}_k\}\right)d^N\mathbf{q}_r^{s'}\right. \nonumber \\ & \left. -\int w^{s's}\prod_{i,j}\delta\left(\left|\mathbf{q}^s_i-\mathbf{q}^s_j\right|-r_{ij}^{s's}\right) \right. \nonumber \\
  & \left. \quad \times P^s_{RC}(\mathbf{Q}^s)\rho^s_{RC}(\{\mathbf{q}^s_k\})d^N\mathbf{q}^s_k\right]
 \end{align}
 
 The evolution equations (\ref{master-closed}) are nonlocal: because of the integral terms, the rate of change of the probability distribution depends on the probability density in the vicinity of a given point. This nonlocality is, however, short-ranged. Indeed, the RC shifts $\mathbf{d}^{ss'}$ are of a molecular size, but spatial distribution of enzymes would vary only a little along such a length. We can use this to approximately convert equations (\ref{master-closed}) to local equations.
 
 To do this, we use an approximate expansion
\begin{align}
 P^{s'}_{RC}(\mathbf{Q}^{s}-\mathbf{d}^{ss'})\approx & P^{s'}_{RC}(\mathbf{Q}^{s})- d^{ss'}_\alpha\frac{\partial P^{s'}_{RC}}{\partial Q^{s'}_\alpha} \nonumber \\
 & + \frac{1}{2}d^{ss'}_\alpha d^{ss'}_\beta \frac{\partial ^2P^{s'}_{RC}}{\partial Q^{s'}_\alpha\partial Q^{s'}_\beta}.
\end{align}
where the derivatives are taken at $\mathbf{Q}^{s'}=\mathbf{Q}^s$.

After substitution of this expansion into equation (\ref{master-closed}), we obtain a set of coupled local evolution equations for RC probability distributions in different chemical states, i.e.
\begin{align}
& \frac{\partial P^s_{RC}}{\partial t}=D^s\frac{\partial^2 P^s_{RC}}{(\partial Q^s_\alpha)^2} -\sum_{s'} A^{ss'}_\alpha \frac{\partial P^{s'}_{RC}}{\partial Q^{s'}_\alpha} \nonumber \\
& +\frac{1}{2}\sum_{s'}B^{ss'}_{\alpha\beta} \frac{\partial ^2P^{s'}_{RC}}{\partial Q^{s'}_\alpha\partial Q^{s'}_\beta} \nonumber \\
 &+\sum_{s'} \left[  \int w^{ss'}\prod_{i,j}\delta\left(\left|\mathbf{q}^{s'}_i-\mathbf{q}^{s'}_j\right|-r_{ij}^{ss'}\right) \right. \nonumber \\
 & \quad \left. \times P^{s'}_{RC}(\mathbf{Q}^s)\rho^{s'}_{RC}(\{\mathbf{q}^{s'}_k\})d^N\mathbf{q}_k^{s'} \right. \nonumber \\ & \left.-\int w^{s's}\prod_{i,j}\delta\left(\left|\mathbf{q}^s_i-\mathbf{q}^s_j\right|-r_{ij}^{s's}\right) \right. \nonumber \\
 & \quad \left. \times P^s_{RC}(\mathbf{Q}^s)\rho^s_{RC}(\{\mathbf{q}^s_k\})d^N\mathbf{q}^s_k\right]
 \end{align}
 where
 \begin{align}
    A^{ss'}_\alpha=& w^{ss'}\int d^{ss'}_\alpha\rho^{s'}_{RC}(\{\mathbf{q}^{s'}_k\}) \nonumber \\
    & \times \prod_{i,j}\delta\Big(|\mathbf{q}^{s'}_i-\mathbf{q}^{s'}_j|-r_{ij}^{ss'}\Big)d^N\mathbf{q}^{s'}_k,
 \end{align}
 \begin{align}
    B^{ss'}_{\alpha\beta}=& w^{ss'}\int d^{ss'}_\alpha d^{ss'}_\beta\rho^{s'}_{RC}(\{\mathbf{q}^{s'}_k\}) \nonumber \\
     & \times \prod_{i,j}\delta\left(\left|\mathbf{q}^{s'}_i-\mathbf{q}^{s'}_j\right|-r_{ij}^{ss'}\right)d^N\mathbf{q}^{s'}_k
 \end{align}
 and $\mathbf{d}^{ss'}=\mathbf{d}^{ss'}(\{\mathbf{q}^{s'}_k\})$ is given by equation (\ref{change}). 
 
 It is convenient to introduce effective rate constants in the steady state
 \begin{align}\label{kappa}
    \kappa^{ss'}= w^{ss'}\int \rho^{s'}_{RC}(\{\mathbf{q}^{s'}_k\})\prod_{i,j}\delta\left(\left|\mathbf{q}^{s'}_i-\mathbf{q}^{s'}_j\right|-r_{ij}^{ss'}\right)d^N\mathbf{q}^{s'}_k.
 \end{align}
 It can be moreover noted that, in all transition conformations differing only by the orientation, the length $d^{ss'}$ of the RC shift is the same and only the directions of the vector $\mathbf{d}^{ss'}$ are different. 
 
 We can write $\mathbf{d}^{ss'}=d^{ss'}\mathbf{n}$ where $\mathbf{n}$ is a unit orientation vector. Because of the statistical isotropy of the equilibrium state, all directions $\mathbf{n}$ are equally probable. Therefore, we obtain
 \begin{align}
    A^{ss'}_\alpha = \frac{1}{4\pi}\kappa^{ss'}d^{ss'}\int n_\alpha d\mathbf{n}=0
 \end{align}
 and
 \begin{align}
    B^{ss'}_{\alpha\beta}= \frac{1}{4\pi}\kappa^{ss'}\left(d^{ss'}\right)^2\int n_\alpha n_\beta d\mathbf{n}=\frac{1}{3}\delta_{\alpha\beta}\kappa^{ss'}\left(d^{ss'}\right)^2
 \end{align}
 where $\int (\dots) d\mathbf{n}$ means integration over all possible orientations of the vector $\mathbf{n}$. Note that a three-dimensional system is considered.
 
 Hence, a set of coupled evolution equations for RC probability distributions in different chemical states becomes derived: 
 \begin{align}\label{effective}
 \frac{\partial P^s_{RC}}{\partial t}=&D^s\frac{\partial^2 P^s_{RC}}{(\partial Q^s_\alpha)^2} +\frac{1}{6}\sum_{s'}\kappa^{ss'}\left(d^{ss'}\right)^2\left.\frac{\partial ^2P^{s'}_{RC}}{(\partial Q^{s'}_\alpha)^2}\right|_{\mathbf{Q}^{s'}=\mathbf{Q}^s} \nonumber \\
 &+\sum_{s'}\left[ \kappa^{ss'}  P^{s'}_{RC}(\mathbf{Q}^{s'}=\mathbf{Q}^s)- \kappa^{s's}P^s_{RC}(\mathbf{Q}^s)\right].
 \end{align}
 The first term in the equation for the probability distribution in chemical state $s$ describes free diffusion with the diffusion coefficient corresponding to this state. The last term accounts for relaxation to the steady state. The middle term is due to the molecular dancing effect.
 
  \subsection{Diffusion enhancement} Using equations (\ref{effective}), the effect of boosted diffusion can be analyzed. Note that, in these equations, probabilities $P^s_{RC}$ in chemical states $s$ depend on different variables $\mathbf{Q}^s$. However, according to the definition of such variables, the difference between them is of the order of the molecular size. But, for diffusion processes, concentration distributions would not significantly vary within such a length.
  
  Hence, we can neglect the difference and treat all of them as a single variable that specifies the position of a molecule on a long length scale. Note, furthermore, that, on such long scales, the difference between RC and CM also disappears. Replacing $\mathbf{Q}^s$ by $\mathbf{R}$, we approximately have 
  \begin{align}\label{states}
 \frac{\partial P^s}{\partial t}=&D^s\frac{\partial^2 P^s}{\partial \mathbf{R}^2} +\frac{1}{6}\sum_{s'}\kappa^{ss'} \left(d^{ss'}\right)^2\frac{\partial^2 P^s}{\partial \mathbf{R}^2} \nonumber \\
 & +\sum_{s'}\left(\kappa^{ss'}  P^{s'}- \kappa^{s's}P^s\right).
 \end{align}
  
  These equations describe both diffusion and relaxation to the chemical steady state. The latter process, with the characteristic timescale of the turnover time, is however much faster than diffusion. Therefore, we can apply the adiabatic approximation and assume that, locally, the system is in the steady state. This means that 
  \begin{align}
      P^s(\mathbf{R},t)=\bar{\pi}^s P(\mathbf{R},t).
  \end{align}
  Here $\bar{\pi}^s$ are steady-state occupation probabilities for different chemical states, such that
  \begin{align}
      \sum_{s'}\left(\kappa^{ss'}\bar{\pi}^{s'}- \kappa^{s's}\bar{\pi}^s\right)=0
  \end{align}
  and
  \begin{align}
      \sum_s\pi^s=1.
  \end{align}
  Substituting this into equations (\ref{states}) and summing them over $s$, an effective diffusion equation for concentration $c=N_0P$ of enzymes is obtained, 
  \begin{align}
    \frac{\partial c}{\partial t}=D_{\mathrm{eff}}\frac{\partial^2 c}{\partial \mathbf{R}^2}  
  \end{align}
  where
  \begin{align}\label{effdiff}
      D_{\mathrm{eff}}=\sum_sD^s\bar{\pi}^s+ \frac{1}{6}\sum_{s,s'}k^{ss'}\left(d^{ss'}\right)^2
  \end{align}
  and 
  \begin{align}\label{rateseff}
     k^{ss'}=\kappa^{ss'}\bar{\pi}^{s'} 
  \end{align}
  are reaction rates in the steady state.
  
  To consider diffusion enhancement, we should subtract the diffusion constant $D_\mathrm{eq}$ in the equilibrium state from $D_\mathrm{eff}$. Some caution is however needed when choosing such ``equilibrium state''. If the considered reaction is irreversible, this is obviously the state in absence of the reaction. In the (more general) case of reversible reactions, this should be the state of chemical equilibrium. In this latter state, reached at a special choice of substrate concentrations, the reactions still go on, but with the forward reactions statistically balanced by the reverse ones. Generally, the equilibrium diffusion constant is 
  \begin{align}
     D_{eq}=\sum_s \bar{\pi}_\textrm{eq}^s D^s+\frac{1}{6}\sum_{s,s'}k_\textrm{eq}^{ss'}\left(d^{ss'}\right)^2
 \end{align}
where $\bar{\pi}_\textrm{eq}^s$ and $k_\textrm{eq}^{ss'}$ are occupation probabilities of different states and rates of transitions between the states at (full chemical) equilibrium. 

Hence, the non-equilibrium change $\Delta D=D_\textrm{eff}-D_\textrm{eq}$ in the diffusion constant is
\begin{align}\label{deltaD2}
     \Delta D=\sum_s \left(\bar{\pi}^s-\bar{\pi}_\textrm{eq}^s\right) D^s+\frac{1}{6}\sum_{s,s'}\left(k^{ss'}-k_\textrm{eq}^{ss'}\right)\left(d^{ss'}\right)^2.
 \end{align}
It consists of two parts. First, as a result of a non-equilibrium reaction, occupation of different states can change and new states can become occupied. The shapes of a molecule are generally different in different states $s$ and, therefore, its diffusion coefficient $D^s$ depends on the chemical state. Thus, when occupation probabilities $\bar{\pi}^s$ are modified, this also affects the (mean) diffusion constant of the enzyme.

The second term in equation (\ref{deltaD2}) represents a contribution from the effects of molecular dancing described by us. In contrast to the first term, it depends not on the occupation numbers of various chemical states, but on the rates $k^{ss'}$ of transitions between them. It also depends on the magnitudes of the shifts $\mathbf{d}^{ss'}$ of the rotation center in these transitions.

As a simple illustration, we can consider an example of an irreversible reaction cycle 
\begin{align} \label{cycle}
    s_0\rightarrow s_1\rightarrow s_2\rightarrow\dots \rightarrow s_K\rightarrow s_0.
\end{align}
Here, $s_0$ represents the ligand-free enzyme and the first transition corresponds to binding of a substrate in this state. The last transition corresponds to product release, returning the enzyme to the initial state. In the steady state of this reaction, the rates of all elementary reactions are the same; they are equal to the enzyme turnover rate $k$. Taking into account that $\bar{\pi}^0=1-\sum_{m=1,K}\bar{\pi}^m$ and that, in absence of the reaction, the enzyme can only be in the state $s_0$, we get
\begin{align}\label{DeltaDmod}
    \Delta D=\sum_{m=1}^K\left(D^s-D^0\right)\bar{\pi}^m+\frac{k}{6}\sum_{m=0}^K\left(d^{m+1,m}\right)^2
\end{align}
where the notation $d^{K+1,K}=d^{0,K}$ is used.

If all reaction steps inside the cycle are very fast (as compared to substrate binding), occupation probabilities of the states inside the cycle are small (i.e., $\bar{\pi}^m\ll1$). For such fast enzymes, the dominant contribution to $\Delta D$ comes from the last molecular dancing term, which is determined solely by the reaction rate.

The excitable dumbbell model of an enzyme, that has been introduced in Section~\ref{sec:discussion}, is described by the reaction cycle (\ref{cycle}) with three states, such that $d^{10}=d^{21}=d$ and $d^{02}=0$. Because the transitions inside the cycle are fast, we approximately have $\Delta D=(1/3)kd^2$ according to equation (\ref{DeltaDmod}). This coincides with the result (\ref{enhancement}) in the limit when rotational diffusion is fast (so that orientational equilibration always takes place between the next reaction steps). 

One can also consider diffusion enhancement in the model described by the reaction cycle (\ref{cycle}) with two states $s_0$ and $s_1$. In this case, we have 
\begin{align}
    \Delta D=\left(D^1-D^0\right)\bar{\pi}^1+\frac{1}{3}kd^2
\end{align}
where $d^{01}=d^{10}=d$ and $k$ is the turnover rate.

\subsection{Systems with spatial gradients}
So far, it has been assumed that a system is uniform. Now, we will extend the theory to systems with spatial gradients. Such gradients typically arise when substrate concentration varies in space. They may however also be caused by variation  of local reaction parameters, such as illumination or pH. Moreover, viscosity of a solution can be also non-uniform, resulting in mobility variation.

Since the system is non-uniform, chemical transition rates (such as, e.g., the rate of substrate binding) depend on the position of the molecule. We shall consider only very slow spatial variations of such rates, so that they do not change significantly on single-molecule length scales. Then, it is not important which point within a molecule is chosen to specify the rates. Therefore, for convenience, we assume below that reaction rates are determined by the RC position before the transition, i.e. $w^{ss'}=w^{ss'}(\mathbf{Q}^{s'})$.

In the non-uniform case, master equations (\ref{master-rot}) are replaced by
\begin{align}
& \frac{\partial p^s_{RC}}{\partial t}=-\frac{\partial H^s_{\alpha}}{\partial Q^s_{\alpha}}-\sum_i\frac{\partial h^s_{i,\alpha}}{\partial q^s_{i,\alpha}} \nonumber \\ 
& + \sum_{s'} \left[ \int  w^{ss'}(\mathbf{Q}^s-\mathbf{d}^{ss'})\prod_{i,j}\delta\left(\left|\mathbf{q}^{s'}_i-\mathbf{q}^{s'}_j\right|-r_{ij}^{ss'}\right) \right. \nonumber \\ & \left. \times \prod_k \delta\left(\mathbf{q}^s_k-\mathbf{q}_k^{s'}+\mathbf{d}^{ss'}\right)p^{s'}_{RC}\left(\mathbf{Q}^s-\mathbf{d}^{ss'},\{\mathbf{q}^{s'}_k\}\right)d^N\mathbf{q}^{s'}_k \right. \nonumber \\ & \left.-w^{s's}(\mathbf{Q}^s)\prod_{i,j}\delta\left(\left|\mathbf{q}^s_i-\mathbf{q}^s_j\right|-r_{ij}^{s's}\right)p^s_{RC}\left(\mathbf{Q}^s,\{\mathbf{q}^s_k\}\right)\right].
 \end{align}
 
 Now, the local steady state depends on the spatial position, so that the steady-state occupation probabilities  of chemical states are  $\bar{\pi}^s=\bar{\pi}^s(\mathbf{Q}^s)$. However, the equilibrium conformational distribution $\rho^s_{RC}(\{\mathbf{q}^s_k\})$ is not affected by the variation of kinetic coefficients; it is still given by the Boltzmann distribution (\ref{Boltzmann}). Importantly, this distribution remains isotropic despite the presence of reaction parameter gradients. The factorization (\ref{factor}) continues to hold.
 
 By repeating the above derivation steps, the following set of coupled evolution equations for probability distributions in different chemical states is obtained:
\begin{align}\label{anti}
 \frac{\partial P^s_{RC}}{\partial t}=& \frac{\partial }{\partial Q^s_\alpha}\left[D^s\frac{\partial P^s_{RC}}{\partial Q^s_\alpha}\right] \nonumber \\
  & +\frac{1}{6}\sum_{s'} \left(d^{ss'}\right)^2\left.\frac{\partial ^2(\kappa^{ss'}P^{s'}_{RC})}{(\partial Q^{s'}_\alpha)^2}\right|_{\mathbf{Q}^{s'}=\mathbf{Q}^s} \nonumber \\
 &+\sum_{s'}\left[   \kappa^{ss'}(\mathbf{Q}^{s'}=\mathbf{Q}^s)  P^{s'}_{RC}(\mathbf{Q}^{s'}=\mathbf{Q}^s) \right. \nonumber\\
 & \quad \left. - \kappa^{s's}(\mathbf{Q}^s)P^s_{RC}(\mathbf{Q}^s)\right]
 \end{align}
where the local reaction rate constants are 
\begin{align}
    \kappa^{ss'}\left(\mathbf{Q}^{s'} \right)=& w^{ss'}\left(\mathbf{Q}^{s'}\right)\int \rho^{s'}_{RC}\left(\{\mathbf{q}^{s'}_k\}\right) \nonumber \\
    & \times \prod_{i,j}\delta\left(\left|\mathbf{q}^{s'}_i-\mathbf{q}^{s'}_j\right|-r_{ij}^{ss'}\right)d^N\mathbf{q}^{s'}_k.
 \end{align}
 The last term in equations (\ref{anti}) describes relaxation to the local steady state. It vanishes for $P^s_{RC}(\mathbf{Q}^s)\propto \bar{\pi}^s(\mathbf{Q}^s)$.

On the timescales longer than chemical relaxation time and on the length scales larger than the characteristic molecular size, the difference between RC and CM distributions is negligible. By taking $P^s_{RC}(\mathbf{Q},t)^s\approx N_0\bar{\pi}^s(\mathbf{R}^s)c(\mathbf{R},t)$ and summing over $s$, an approximate eolution equation for the local concentration $c$ of enzymes in systems with spatial gradients is obtained 
\begin{align} \label{antii}
    \frac{\partial c}{\partial t}=&\frac{\partial}{\partial \mathbf{R}}\left[\sum_s D^s\frac{\partial }{\partial \mathbf{R}}(\bar{\pi}^s c)\right] \nonumber \\
    &+\frac{1}{6}\sum_{s,s'}\left(d^{ss'}\right)^2\frac{\partial^2}{\partial \mathbf{R}^2}\left[\kappa^{ss'}\bar{\pi}^{s'}c\right].
\end{align}

\end{document}